\documentstyle[aps,pre,subeqnarray,twocolumn,epsfig]{revtex}

\begin{document}

\draft

\preprint{20. May. 2001}
\title{Microscopic dynamics of molecular liquids and glasses: Role of
orientations and translation-rotation coupling}
\author{T.~Theenhaus, R.~Schilling}
\address{Institut f\"ur Physik, Johannes Gutenberg--Universit\"at, 
Staudinger Weg 7,
D--55099 Mainz, Germany}
\author{A. Latz \footnote{author to whom
correspondence should be addressed}} 
\address{  
Institut f\"ur Physik, Reichenhainer Strasse 70, TU - Chemnitz,
D-09107 Chemnitz, Germany}
 \author{M. Letz}
\address{Schott Glas,
Research and Developement, Hattenbergstr. 10, 55014 Mainz, Germany}
\maketitle

\begin{abstract}
We investigate the dynamics of a fluid of dipolar hard spheres in its liquid
and glassy phase, with emphasis on the microscopic time or frequency regime.
This system shows rather different glass transition scenarios related to its
rich equilibrium behavior which ranges from a simple hard sphere fluid to a
long range ferroelectric orientational order. In the liquid phase close to
the ideal glass transition line and in the glassy regime a medium range
orientational order occurs leading to a softening of an orientational mode.
To investigate the role of this mode we use the molecular mode-coupling
equations to calculate the spectra $\phi _{lm}^{\prime \prime }(q,\omega )$
and $\chi _{lm}^{\prime \prime }(q,\omega )$. In the center of mass spectra $
\phi _{00}^{\prime \prime }(q,\omega )$ and $\chi _{00}^{\prime \prime
}(q,\omega )$ we found besides a high frequency peak at $\omega _{hf}$ a
peak at $\omega _{op}$, about one decade below $\omega _{hf}$. $\omega _{op}$
has almost no $q$-dependence and exhibits an ``isotope'' effect $\omega
_{op}\propto I^{-1/2}$, with $I$ the moment of inertia. We give evidence
that the existence of this peak is related to the occurrence of the medium
ranged orientational order. It is shown that some of these feature also
exist for schematic mode coupling models.
\end{abstract}


\section{Introduction}

\label{cap1} The dynamical properties of liquids and glasses still are a
challenging problem. In the vicinity of the glass transition the frequency
(or time) range can be decomposed into at least three different domains, the 
$\alpha $, $\beta $ and microscopic regime. The first one describes the
structural relaxation which dramatically slows down when the glass
transition is approached from above. In the idealized mode-coupling theory
(MCT) for simple liquids \cite{goetze91,goetze92,schilling94,cummins99} and
for molecular systems \cite{franosch97a,schilling97,fabbian99} it even stops at a
critical temperature $T_{c}$. Probably the most interesting result of MCT is
the existence of the so-called $\beta $-relaxation which describes the
dynamics within a cage of particles above and below $T_{c}$. The
corresponding $\beta $-frequency scale is much larger than that for the $
\alpha $-relaxation. At still higher frequencies there are vibrational and
librational motion which constitute the microscopic regime.

One may say that most of the attention in the field of glassy dynamics
during the last fifteen years has been devoted to the $\alpha $- and $\beta $
-relaxation. This activity has been mainly stimulated by MCT which has
predicted in these two regimes scaling laws with a diverging $\alpha $- and $
\beta $-time scale. These predictions have been tested intensively by
experiments and numerical simulations. A satisfactory agreement has been
found for many glass forming systems
\cite{yip95,kob99,goetze99b,tao91,theis98,schilling00,winkler00,fabbian99,theis00}.
Experimental and simulational results do not exhibit any singular or
crossover behavior for microscopic frequencies $\omega $, i.e. for $\omega
\geq 1$ THZ. Nevertheless in that regime an interesting phenomenon occurs in
most glasses but not in crystals and
colloidal glasses, which is the so called \emph{boson-peak}. Indications for
this peak came from two different sides. First, the low temperature specific
heat $c(T)/c_{D}(T)$ scaled by the phonon contribution $c_{D}(T)\propto
T^{3} $ shows a peak at about 10 K for several glass formers (see
e.g. \cite{pohl81}).  This excess with respect to $c_{D}(T)$ around 10 K
implies the 
existence of additional excitations besides the long wavelength acoustic
phonons. Second, Raman spectra $I(\omega )$ compared to the phonon
contribution $I_{D}(\omega )\propto \omega ^{2}$ exhibit at about 1 THZ an
excess, as well (see e.g. \cite{malinovski86}). Since the temperature
dependence of the excess intensity scales with the Bose-distribution
function $n_{B}(T)$, the peak is called boson-peak.

That these two observations might have a common origin was first shown by
Buchenau et al. \cite{buchenau84,buchenau86}. For vitreous silica these
authors also found an excess with respect to $g_{D}(\omega )\propto \omega
^{2}$ for the vibrational density of states $g(\omega )$ determined from
inelastic neutron scattering (INS) data. Using $g(\omega )$ to calculate $
c(T)/c_{D}(T)$ led to a good agreement with the result from heat capacity
measurements. Particularly, the peak positions for both results coincided.
Although the boson-peak does not seem to posses any singular $\omega $- or $T
$-dependence it is a \emph{universal} phenomenon for all systems with 
\textit{exclusion} of glass forming colloids, in the sense that it appears
more or less for almost all glass formers. For, e.g. LiCl solutions 
\cite{tao91} and Orthoterphenyl \cite{cummins97}, it has been stressed that
the 
boson-peak, together with the narrowing of a central peak at $\omega =0$,
develops continously from the liquid to the glassy phase.

Despite considerable experimental and numerical efforts its microscopic
origin is still not satisfactorily understood. Two reasons might be
responsible for that. First, e.g.~in case of light scattering the \emph{
precise} connection between the measured quantity and the basic theoretical
objects, the time- or frequency-dependent site-site or molecular correlation
functions for molecular liquids, is not known. For instance, it has been
shown that several coupling mechanism between light and distinct modes of
liquid ZnCl$_{2}$ exist which have different $\omega $-dependent coupling
constants \cite{ribeiro98}. This may complicate the determination of $
g(\omega )$ from Raman spectra. Second, the relationship between the various 
$t$- or $\omega $- dependent correlators and microscopic modes obtained from
a diagonalization of the dynamical matrix is not obvious. In addition, these
correlators can only be calculated analytically under serious
approximations. Apparently, numerical investigations represent a powerful
tool since their microscopic nature allows to calculate both, the
correlators or spectra and under certain conditions the microscopic modes.
The possible character of these modes range between:\newline
\vspace{.5cm}

\begin{centering}
\centerline{\begin{tabular}{rcl}
vibrational & $\longleftrightarrow$ & relaxational\\
extended & $\longleftrightarrow$ & localized\\
propagating & $\longleftrightarrow$ & non-propagating\\
harmonic & $\longleftrightarrow$ & anharmonic\\
longitudinal & $\longleftrightarrow$ & transversal\\
coherent & $\longleftrightarrow$ & random\\
translational & $\longleftrightarrow$ & rotational\\
acoustic & $\longleftrightarrow$ & optic\\
\end{tabular}}
\end{centering}

\vspace{.5 cm}

\noindent Since these feature can occur in combinations, the complexity of
the problem becomes obvious. In addition, different experimental approaches
to a specific material or a specific measurement of different type of glass
formers, e.g. strong or fragile ones, may exhibit different features of the
same phenomenon.

Without demanding completeness let us shortly review the present status.
Inelastic X-ray scattering (IXS) on v-SiO$_2$ gave evidence that \emph{
propagating} acoustic sound waves exist even above $\omega_{BP}$, the
position of the boson-peak \cite{benassi96}. This has let these authors to
conclude that the propagating modes are also involved in the boson-peak
itself. On the other side a crossover at $\omega_{BP}$ from \emph{propagating
} to \emph{localized} (strongly scattered) acoustic modes was deduced for
v-SiO$_2$ from INS and IXS \cite{foret96}. Comments on this controversy are
given in Ref. \cite{foret97,benassi97,matic01}. However, interpretation of
IXS-experiments which are free of any model have recently strengthened at
least the fact that there are propagating modes above $\omega_{BP}$ 
\cite{masciovecchio00,pilla00}. MD-simulations for v-SiO$_2$ have given 
evidence
that the boson-peak modes can \emph{not strongly be localized} and that
there is a contribution of \emph{transverse propagating} modes in the
boson-peak regime \cite{horbach99,horbach99a}. IXS for LiCl-solutions and
glycerol seem to detect propagating, \emph{longitudinal} modes at
$\omega_{BP}$ \cite{masciovecchio96}, whereas a MD-simulation for H$_2$O
shows a mixing of propagating \emph{longitudinal} and \emph{transversal}
modes, at least for $q$ large enough ($q>4nm^{-1}$) \cite{sampoli97}.

Further important information about the nature of the boson-peak modes comes
from a normal mode analysis for a system of soft spheres \cite{schober96} and for SiO$
_{2}$ deep in its glass phase 
\cite{taraskin97a,taraskin97b,taraskin99a,taraskin99b}. There it has been found
that these modes are \emph{harmonic} and \emph{quasi localized} 
\cite{schober96,taraskin99a}
. They occur due to \emph{hybridization} of localized low frequency optic
modes with propagating acoustic states \cite{taraskin97a} and are \emph{
extended} but \emph{non-propagating}. In addition each normal mode has a 
\emph{coherent } and a \emph{random} component. The latter finding is
consistent with other numerical results for a Lennard-Jones liquid 
\cite{mazzacurati96}, liquid ZnCl$_{2}$ \cite{ribeiro98,foley95,ribeiro99} and
conclusions drawn from experimental data for various glass formers 
\cite{engberg98,sokolov99}. Since ZnCl$_{2}$ has been considered in its liquid
phase an instantaneous normal mode analysis is feasible, only. Restriction
to the stable modes accurately reproduces in the microscopic frequency
regime the dynamical structure factor $S(q,\omega )$ obtained from a
MD-simulation \cite{ribeiro99}. This implies a harmonic mode character at
temperatures even above the MCT glass-transition temperature $T_{c}$.

There is not much analytical work on the boson peak. For a \emph{harmonic}
crystal with \emph{random} spring constants it has been shown
\cite{schirmacher89,schirmacher98} 
that an excess density of states follows. It has recently been stressed 
\cite{martin00} that the approach in \cite{schirmacher98} has some shortcomings
which may be removed by use of \emph{off-lattice} models \cite{mezard99}.
Another theoretical framework is the soft potential model \cite
{karpov83,klinger88,galperin85,buchenau92} which allows to describe the low
temperature 
anomalies below 1K and those at 10-20 K, which includes the excess density of
states.

As already mentioned above, MCT has been rather successful in describing the 
$\alpha $- and $\beta $-dynamics. Schematic MCT-models \cite{goetze91} where
the wave number dependence has been neglected were also used to describe
experimental spectra in the microscopic regime, including the boson peak 
\cite{tao91,simionesco95,goetze96,franosch97c,das99,ruffle99,goetze00b}. 
Whereas the results in
Ref. \cite{tao91} (cf. Fig. 6) and Ref. \cite{franosch97c} (cf. Fig. 4) yield
spectra which are reminiscent to boson peak spectra the line shape does not
come out, satisfactorily. However, recently a detailed MCT-investigation for
a glass of hard spheres was performed, including the $q$- dependence \cite
{goetze00}. Besides a high frequency peak at $\omega _{HF}$ an additional
peak at $\omega _{AOP}$, about one decade below $\omega _{HF}$, has been seen
for 
a volume fraction $\varphi =0.6$. This peak, which strongly resembles that
in Refs. \cite{tao91} (Fig. 6) and \cite{goetze96} (Fig.~\ref{fig7n}), originates
from the distribution of harmonic oscillators within the cages and has been
called \emph{anomalous oscillation peak} (AOP). As explicitely demonstrated 
\cite{goetze00} it shares many features with the boson peak.

In contrast to a system of hard spheres or binary van der Waals liquids,
molecular liquids also have orientational degrees of freedom. The question
which one may ask is: Does the boson peak also involve orientational motion?
Indeed one of its first interpretations for v-SiO$_{2}$ were coupled
rotations of SiO$_{4}$-tetrahedra \cite{buchenau84}. Such an interpretation
has been supported by MD-simulations for v-SiO$_{2}$ \cite
{taraskin97b,guillot97} and ZnCl$_{2}$ \cite{ribeiro99} and  by
neutron scattering experiments, proving non-sound-like contributions around $
\omega _{BP}$ \cite{wischnewski98}. 
Dielectric loss measurements probe the orientational dynamics, only. 
Since these measurements, e.g. for Glycerol and Propylene Carbonate, 
also exhibit a
boson peak \cite{schneider99,lunkenheimer00con} gives additional
evidence 
that this peak may also be related to the orientational degrees of freedom. 
The role of orientational modes becomes
even more clear from the experiments on ethanol \cite{ramos97,fischer99}.
Around 100 K, Ethanol can occur in several phases: glass phase,
orientational glass phase, crystalline phase and a rotator phase. The center
of mass positions of the molecules are frozen in an amorphous structure for
the glass phase, and in a crystalline structure for the three other phases.
The orientational dynamic is non-ergodic for the glass and the orientational
glass and ergodic for the crystal and the rotator phase. INS has shown the
existence of a boson peak for the orientational glass which does not differ
much from that in the glass. Even in the rotator phase there is a boson
peak, however, shifted to lower frequencies. Similarly to structural glasses
the orientational glass phase also exhibits an excess in the specific heat $
c(T)/c_{D}(T)$ \cite{ramos97}. These findings suggest that these excess modes
are primarily related to the orientational degrees of freedom. 

In the
present paper we will not calculate the density of states, but the
susceptibility and correlation spectra of the collective dynamics. These
quantities show the appearance of an extra peak about a decade below the
high frequency peak, where the former also originates from the orientational
degrees of motion.

The outline of the paper is as follows. In Sec. \ref{cap2}, we will shortly
review the derivation of the equation of motion for the relevant correlators
of rigid, linear molecules and the mode-coupling approximation. In addition,
we will discuss the linearized equations. The molecular mode-coupling
equations are solved numerically for a liquid of dipolar hard spheres in
Sec. \ref{cap3.2}. It is shown in Sec. \ref{cap4} that the qualitative
features of the susceptibility spectra obtained for dipolar hard spheres can
be derived from a schematic model. Finally, Section \ref{cap5} contains a
discussion of the results and some conclusions.

\section{EQUATIONS OF MOTION}

\label{cap2} In the first part of this section we will present the equations
of motion for the most relevant correlation functions of a molecular liquid.
The second part contains a discussion of the corresponding linearized
equations, which yields information on the microscopic time scale and in the
third part we will shortly review the mode-coupling approximation.

\subsection{Mori equation}

\label{cap2.1} We restrict ourselves to a system of $N$ rigid and \textit{
linear} molecules with mass $M$ and moment of inertia $I$. There are two
possibilities to describe molecular liquids: a \emph{site-site} or a \emph{
molecular} representation \cite{hansen86}. The latter, which will be chosen
here, decomposes the 5$N$ degrees of freedom into 3$N$ translational and 2$N$
orientational ones. The starting point is the microscopic tensorial
one-particle density mode 
\begin{equation}
\rho _{lm}(\vec{q},t)=\sqrt{4\pi }\,\,i^{l}\,\sum_{n=1}^{N}e^{i\,\vec{q}
\cdot \vec{x}_{n}(t)}\,Y_{lm}(\Omega _{n}(t))  \label{eq1}
\end{equation}
and the corresponding translational and rotational current density mode 
\begin{equation}
\vec{j}_{lm}^{\alpha }(\vec{q},t)=\sqrt{4\pi }\,\,i^{l}\,\sum_{n=1}^{N}\vec{v
}_{n}^{\alpha }(t)e^{i\,\vec{q}\cdot \vec{x}_{n}(t)}\,Y_{lm}(\Omega _{n}(t))
\label{eq2}
\end{equation}
for $\alpha =T$ and $R$, respectively. $\vec{x}_{n}(t)$ and $\Omega
_{n}(t)=(\theta _{n}(t),\varphi _{n}(t))$ are the center of mass position
and the orientation (specified by the two angles $\theta _{n}$ and $\varphi
_{n}$) of the $n$-th molecule at time $t$, respectively. $Y_{lm}(\theta
,\varphi )$ are the spherical harmonics with $l=0,1,2,...$, $-l\leq m\leq l$
and the velocities $\vec{v}_{n}^{\alpha }(t)$ are defined by

\begin{equation}  \label{eq3}
\vec{v}^{\alpha}_{n}(t) = \left\{ 
\begin{array}{l@{\quad ,\quad}l}
\vec{\dot{x}}_n(t) & \alpha=T \\ 
\vec{\omega}_n(t) & \alpha=R
\end{array}
\right .
\end{equation}
where $\vec{\omega}_n(t)$ is the corresponding angular velocity. $\rho_{lm}$
and $\vec{j}^\alpha_{lm}$ are related to each other by the continuity
equation: 
\begin{equation}  \label{eq4}
\dot{\rho}_{lm}(\vec{q},t)=i \!\! \sum _{\alpha = T,R} q_l^\alpha(\vec{q}
)\,j_{lm}^\alpha(\vec{q},t)
\end{equation}
with: 
\begin{subeqnarray}
j_{lm}^T(\vec{q},t)=\frac{1}{q} \vec{q}\cdot \vec{j}_{lm}^T(\vec{q},t) ,\quad
q=|\vec{q}| 
\\
j_{lm}^R(\vec{q},t)=\frac{1}{\sqrt{l(l+1)}} \vec{L}\cdot \vec{j}^R_{lm}(\vec{q},t)
\label{eq5}
\end{subeqnarray}
the ''longitudinal'' translational and rotational current density and 
\begin{equation}  \label{eq6}
q^\alpha _l(\vec{q}) := \left\{ 
\begin{array}{r@{\quad,\quad}l}
q \qquad & \alpha = T, \quad \forall l \\ 
\sqrt{l(l+1)} & \alpha = R, \quad \forall \vec{q} \quad.
\end{array}
\right.
\end{equation}
$\vec{L}$ denotes the operator of angular momentum.

The time- and $(\vec{q};lm,l^{\prime }m^{\prime })$-dependent correlation
functions: 
\begin{equation}
S_{lm,\,l^{\prime }m^{\prime }}(\vec{q},t)=\frac{1}{N}\langle \delta \rho
_{lm}^{\ast }(\vec{q},t)\delta \rho _{l^{\prime }m^{\prime }}(\vec{q}
,0)\rangle   \label{eq7}
\end{equation}
and 
\begin{equation}
J_{lm,l^{\prime }m^{\prime }}^{\alpha \alpha ^{\prime }}(\vec{q},t)=\frac{1}{
N}\langle j_{lm}^{\alpha \ast }(\vec{q},t)\,j_{l^{\prime }m^{\prime
}}^{\alpha ^{\prime }}(\vec{q},0)\rangle   \label{eq8}
\end{equation}
are of particular experimental and theoretical importance, at least for $
l=l^{\prime }\leq 2$. Here, $\delta \rho _{lm}(\vec{q},t)=\rho
_{lm}(\vec{q},t)-\left\langle \rho _{lm}(\vec{q},t)\right\rangle $ is the
corresponding fluctuation of the density. ${\mathbf{S}}(\vec{q}
,t)=(S_{lm,\,l^{\prime }m^{\prime }}(\vec{q},t))$ form a complete set for any
time dependent two-point density correlator. Particularly, the partial
dynamical structure factors in a site-site description are linear
superpositions of the molecular correlators $S_{lm,l^{\prime }m^{\prime }}(
\vec{q},t)$, but not vice versa \cite{theis99}. In addition, we introduce
the density-current density correlator:
\begin{equation}
S_{lm,\,l^{\prime }m^{\prime }}^{\alpha }(\vec{q},t)=\frac{1}{N}\left\langle
j_{lm}^{\alpha \ast }(\vec{q},t)\delta \rho _{l^{\prime }m^{\prime }}(\vec{q}
,0)\right\rangle   \label{eqAs1}
\end{equation}
Mori-Zwanzig formalism has been used to derive equations of motion for
${\mathbf{S}}(
\vec{q},t)$ for a single linear molecule in a liquid of isotropic particles 
\cite{franosch97a} and for a molecular liquid of linear \cite{schilling97}
and arbitrary molecules \cite{fabbian99}. Similar work has been done in a
site-site description \cite{chong98}. A comparison between the tensorial and
the site-site mode coupling theory has recently been performed for a single
dumbell in a liquid of hard spheres \cite{chong00}.

Choosing $\delta \rho _{lm}$ and both scalar current densities $
j_{lm}^{\alpha }~,\alpha =T,R$ as slow variables, the Mori-Zwanzig formalism
leads to the following set of equations:
\begin{subeqnarray}
\dot{{\mathbf{S}}}(\vec{q},t)+i\sum\limits_{\alpha }{\mathbf{q}}
^{\alpha }{\mathbf{S}}^{\alpha }(\vec{q},t)=0 
\\\nonumber 
\dot{\mathbf{S}}^{\alpha }(\vec{q},t)+i{\mathbf{q}}^{\alpha }
{\mathbf{J}}^{\alpha }(\vec{q}){\mathbf{S}}^{-1}\left(\vec{q},0\right) 
{\mathbf{S}}(\vec{q},t)+
\\
{\mathbf{J}}^{\alpha }(\vec{q})\int\limits_{0}^{t}dt^{\prime }\sum\limits_{\alpha
^{^{\prime }}}{\mathbf{m}}^{\alpha \alpha ^{\prime }}\left(\vec{q}, t-t^{\prime
}\right) {\mathbf{S}}^{\alpha ^{\prime }}(\vec{q},t^{\prime })=0  \label{eqAs2}
\end{subeqnarray}

with initial conditions:
\begin{eqnarray}
{\mathbf{S}}(\vec{q},0) &\equiv &{\mathbf{S}}(\vec{q})  \label{eqAs3} \\
\dot{{\mathbf{S}}}(\vec{q},0) &\equiv &0\equiv {\mathbf{S}}
^{\alpha }(\vec{q},0)   \\
\dot{{\mathbf{S}}}^{\alpha }(\vec{q},0) &\equiv &-i\mathbf{J} (\vec{q})
^{\alpha }{\mathbf{q}}^{\alpha }  
\end{eqnarray}

and
\begin{eqnarray}
\left( {\mathbf{q}}^{\alpha }\right) _{lm,l^{\prime }m^{\prime }}
&=&q_{l}^{\alpha }(\vec{q})\delta _{ll^{\prime }}\delta _{mm^{\prime }}  \label{eqAs4}
\\
\left( {\mathbf{J}}^{\alpha }(\vec{q})\right) _{lm,l^{\prime }m^{\prime }}
&\equiv&J_{lm,l^{\prime }m^{\prime }}^{\alpha \alpha ^{\prime }}(\vec{q})
=\frac{k_{B}T}{I^{\alpha }}\delta _{\alpha \alpha ^{\prime }}\delta
_{ll^{\prime }}\delta _{mm^{\prime }}  
\end{eqnarray}

Here, ${\mathbf{S}}(\vec{q})$ and ${\mathbf{J}}^{\alpha \alpha ^{\prime }}\left( 
\vec{q}\right) $ denote, respectively, the static density and current
density correlation matrix and 
\begin{equation}
I_{\alpha }=\left\{ 
\begin{array}{cc}
M, & \alpha =T \\ 
I, & \alpha =R
\end{array}
\right. 
\label{eq10}
\end{equation}

Note first that these equations are completely equivalent to the set of
equations in \cite{schilling97}. However, they are given in a different
representation, which makes its numerical solution convenient. Second, this
set of equations is still exact, but needs an expression for the memory
kernels $m_{lm,l^{\prime }m^{\prime }}^{\alpha \alpha ^{\prime }}\left( \vec{
q},t\right) $. 
This is where approximations come in. Their nature depends strongly on the
physical situation: for example $\mathbf{m}^{\alpha \alpha ^{\prime }}$ for
a supercooled liquid will be quite different from that for a liquid at
higher temperatures. Third, instead of choosing the scalar current densities
one could also use each Cartesian component $j_{lm}^{\alpha i}$, $i=x,y,z$
as slow variables. This has been done recently for $\alpha =T$, but not for $
\alpha =R$ in order to discuss the role of transversal currents on light
scattering spectra \cite{letz00}. The resulting equations again are exact but
involve memory kernels $m_{lm,l^{\prime }m^{\prime }}^{\alpha i,\alpha^{\prime
}i}\left( \vec{q},t\right) $.

\subsection{Linearized equations of motion}

Since the memory kernels will not be independent functions of the tensorial
density correlations $\mathbf{S}\left( \vec{q},t\right) $, the third term in
Eq.~(\ref{eqAs2}b) is a kind of \textit{nonlinearity}. Within MCT $\mathbf{m}
^{\alpha \alpha ^{\prime }}\left( \vec{q},t\right) $ is approximated by
superpositions of products $\mathbf{S}\left( \vec{q}_{1},t\right) \mathbf{S}
\left( \vec{q}_{2},t\right) $ with $\vec{q}=\vec{q}_{1}+\vec{q}_{2}$, which
makes obvious the nonlinear character of the equations of motion. It is this
nonlinearity which leads to a slowing down of the structural relaxation by
decreasing the temperature or increasing the density of a liquid. This
behavior takes place on the liquid side as a two-step relaxation process
characterized by two diverging time scales $t_{\sigma }\sim |T-T_{c}|^{-
\frac{1}{2a}}$ and $\tau \sim \left( T-T_{c}\right) ^{-\gamma }$ where $a$
and $\gamma $ are positive and $T_{c}$ is the ideal glass transition
temperature. The time scale for $t_{\sigma }$ and $\tau $ is determined by a
microscope scale $t_{0}$. $t_{0}$ depends on inertia and damping
effects where the latter are due to the regular part of $\mathbf{m}^{\alpha
\alpha ^{\prime }}\left( \vec{q},t\right) $, accounting for the fast motions.
Because our main interest is the microscopic dynamics, we can
get an estimate of the microscopic time scale by neglecting the memory term
including its regular part. This results in a set of \textit{linear}
equations from which one immediately obtains for the normalized correlator $
\mbox{\boldmath {$\Phi $}}(\vec{q},t)=
\mathbf{S}^{-\frac{1}{2}}(\vec{q})\mathbf{S}(\vec{q},t)\mathbf{
S}^{-\frac{1}{2}}(\vec{q})$:
\begin{equation}
\ddot{\mbox{{\boldmath {$\Phi $}}}}(\vec{q},t)+\mbox{\boldmath {$\Omega $}}
^{2}(\vec{q})\mbox{\boldmath {$\Phi $}}(\vec{q},t)=0  \label{eq13a}
\end{equation}
with initial conditions 
\begin{equation}
\mbox{\boldmath {$\Phi $}}(\vec{q},0)={\mathbf{1}},\qquad \dot{
\mbox{\boldmath{$\Phi $}}}(\vec{q},0)=\mathbf{0}.  \label{eq13b}
\end{equation}
and the hermitian frequency matrix squared:  

\begin{equation}
\mbox{\boldmath {$\Omega $}}^{2}(\vec{q})
={\mathbf{S}}^{-\frac{1}{2}}(\vec{q})\sum_{\alpha
}\left( {\mathbf{q}}^{\alpha }\right) ^{2}{\mathbf{J}}^{\alpha }{\mathbf{S}}^{-
\frac{1}{2}}(\vec{q})  \label{eq17a}
\end{equation}

or with Eq.~(\ref{eq6}), (\ref{eqAs4}) and (\ref{eq10})
\begin{eqnarray}
\left(\mbox{\boldmath {$\Omega $}}^{2}(\vec{q})\right)_{lm.l^{\prime }m^{\prime }}
&=&\sum_{l^{\prime \prime }m^{\prime \prime }}\left( {\mathbf{S}}^{-\frac{1}{2}
}(\vec{q})\right) _{lm,l^{\prime \prime }m^{\prime \prime }}  \label{eq17b}
\\
&&\times
\left[ \frac{k_{B}T}{M}q^{2}+\frac{k_{B}T}{I}l^{\prime \prime }\left(
l^{\prime \prime }+1\right) \right]    \\
&&\times
\left( {\mathbf{S}}^{-\frac{1}{2}}(\vec{q})\right)_{l^{\prime \prime
}m^{\prime \prime },l^{\prime }m^{\prime }}  
\end{eqnarray}

Here, some comments are in order. First, the static correlators $
S_{lm,l^{\prime }m^{\prime }}(\vec{q})$ and therefore $\Omega_{lm,l^{\prime
}m^{\prime }}(\vec{q})$ are not diagonal in $l$ and $l^{\prime }$ in
general. Accordingly, translational and rotational modes generally are
coupled to each other for given $\vec{q}$. Second, this coupling vanishes in
the limit $q\rightarrow 0$, because the static correlators $S_{lm,l^{\prime
}m^{\prime }}(\vec{q})$ and therefore $\Omega _{lm,l^{\prime }m^{\prime }}(
\vec{q})$ (cf. Eq. (\ref{eq17b})) become diagonal and independent from each
other for an
isotropic liquid:
\begin{equation}
\Omega _{lm,l^{\prime }m^{\prime }}(\vec{q})\rightarrow \omega _{l}(q)\delta
_{ll^{\prime }}\delta _{mm^{\prime }}  \label{eqE1}
\end{equation}
with 
\begin{subeqnarray}
\omega_0(q) & = &
\sqrt{\frac{k_BT}{MS_0}} \, q
\\
\omega_l(q) & = &
\sqrt{\frac{k_BT}{IS_l}}\, \sqrt{l(l+1)}, \quad l>0 
\label{eqE2}
\end{subeqnarray}

the translational and rotational frequencies for $q\rightarrow 0$ and $
S_{l}\equiv S_{l0,l0}(\vec{q}=0)$. $\omega _{0}(q)$ describes the well known 
\emph{acoustic} isothermal sound wave dispersion and $\omega_{l}\left(
q\right) $ the \emph{optic} rotational frequencies for $l>0$. 

Third, the limit $q\rightarrow 0$ was already discussed for a molecular
liquid using a \emph{site-site} description \cite{ricci89}. These authors
also have set the corresponding memory matrix to zero, which, by the way, is
completely equivalent to a short time expansion of the equation of motion
in leading order. But there
the coupling between the partial dynamical structure factors does not vanish
for $q\rightarrow 0$. In this respect the choice of a molecular
representation which takes care of the isotropy of the microscopic
Hamiltonian is the most natural one, at least for $q\rightarrow 0$ and $
t\rightarrow 0$.

So far we have considered the \textit{liquid} phase only. The idealized
version of MCT, which neglects so-called hopping processes, is a theory
developed on the liquid side of the glass transition, because it uses as an
input static correlators in equilibrium. It may also be used below, but
close to the transition point where the singular behavior still dominates.
Whether MCT is even capable to describe dynamical feature far below the
glass transition point is unclear. Nevertheless, we will also apply MCT for
parameters deeper in the {\em glass} phase. An ideal glass is a non-ergodic phase
with non-ergodicity parameters (not normalized):
\begin{equation}
F_{\ell m,\ell ^{\prime }m'}(\vec{q})=\lim_{t\rightarrow
\infty }S_{\ell m,\ell ^{\prime }m^{\prime }}(\vec{q},t)  \label{eqAs6}
\end{equation}
which are nonzero. One can easily prove that Eqs.~(\ref{eqAs2}) then imply
that 
\begin{equation}
C_{\ell m,\ell ^{\prime }m^{\prime }}^{\alpha \alpha ^{\prime }}(\vec{q})=
\lim_{t\rightarrow \infty }m_{\ell m,\ell ^{\prime }m^{\prime
}}^{\alpha \alpha ^{\prime }}(\vec{q},t)  \label{eqAs7}
\end{equation}
are nonzero, too, and that
\begin{equation}
\sum\limits_{\alpha ,\alpha ^{\prime }}{\mathbf{q}}^{\alpha }[({\mathbf{C}}
^{\beta \beta ^{\prime }}(\vec{q}))^{-1}]^{\alpha \alpha ^{\prime }}
{\mathbf{q}}
^{\alpha ^{\prime }}
=
({\mathbf{S}}(\vec{q})-{\mathbf{F}}(\vec{q})){\mathbf{F}}^{-1}(
\vec{q}){\mathbf{S}}(\vec{q}).  \label{eqAs8}
\end{equation}
Therefore, we introduce ${\mathbf{\hat{S}}}(\vec{q},t)$ and $\hat{{\mathbf{m}}}
^{\alpha \alpha ^{\prime }}(\vec{q},t)$ such that
\begin{eqnarray}
{\mathbf{S}}(\vec{q},t) &=&{\mathbf{F}}(\vec{q})+{\mathbf{\hat{S}}}(\vec{q},t)
\label{eqAs9} \\
{\mathbf{m}}^{\alpha \alpha ^{\prime }}(\vec{q},t) &=&{\mathbf{C}}^{\alpha
\alpha ^{\prime }}\left( \vec{q}\right) +\hat{{\mathbf{m}}}^{\alpha \alpha
^{\prime }}(\vec{q},t)  
\end{eqnarray}
in analogy to the approach in \cite{goetze00}. Substitution of
Eq.~(\ref{eqAs9}) into Eq.~(\ref{eqAs2}) yields:

\begin{subeqnarray}
\label{eqAs10} 
\dot{\hat{{\mathbf{S}}}}(\vec{q},t)+i\sum\limits_{\alpha }
{\mathbf{q}}^{\alpha }{\mathbf{S}}^{\alpha }(\vec{q},t)=0 
\\\nonumber
\dot{{\mathbf{S}}}^{\alpha }(\vec{q},t)+
i{\mathbf{q}}^{\alpha }{\mathbf{J}}^{\alpha }(\vec{q})
{\mathbf{S}}^{-1}(\vec{q}){\mathbf{\hat{S}}}(\vec{q},t)+
\\\nonumber
{\mathbf{J}}^{\alpha}(\vec{q}) \int\limits_{0}^{t}dt^{\prime }
\sum_{\alpha'} {\mathbf{\hat{m}}}^{\alpha \alpha
^{\prime }}(\vec{q},t-t^{\prime })
{\mathbf{S}}^{\alpha ^{\prime }}(\vec{q},t^{\prime })+  
\\\nonumber
+i{\mathbf{q}}^{\alpha }{\mathbf{J}}^{\alpha } (\vec{q}) 
{\mathbf{S}}^{-1}(\vec{q}){\mathbf{F}}(\vec{q})+
\\
{\mathbf{J}}^{\alpha } (\vec{q})
\sum\limits_{\alpha ^{\prime }}{\mathbf{C}}
^{\alpha \alpha ^{\prime }}(\vec{q}) 
\int\limits_{0}^{t}dt^{\prime } {\mathbf{S}}^{\alpha
^{\prime }}(\vec{q},t^{\prime }) =0
\end{subeqnarray}
with initial conditions:
\begin{equation}
{\mathbf{\hat{S}}}(\vec{q},0)={\mathbf{S}}(\vec{q})-{\mathbf{F}}(\vec{q})
\label{eqAs11}
\end{equation}
and
\begin{equation}
\lim_{t\to \infty }{\mathbf{\hat{S}}}(\vec{q},t)=0,\quad 
\lim_{t\to \infty }{\mathbf{\hat{m}}}^{\alpha \alpha
^{\prime }}(\vec{q},t)=0.  \label{eqAs12}
\end{equation}

With the same argumentation as above we linearize Eqs.~(\ref{eqAs10}b) by taking $
{\mathbf{\hat{m}}}^{\alpha \alpha ^{\prime }}(\vec{q},t)$ to be zero. Then,
taking the time derivative of Eq.~(\ref{eqAs10}b) and using (\ref{eqAs10}a) 
we get for ${\mathbf{\hat{S}}}
^{\alpha }(\vec{q},t)\equiv({\mathbf{J}}^{\alpha
}(\vec{q}))^{-\frac{1}{2}}{\mathbf{S}}^{\alpha }(\vec{q},t)$:
\begin{equation}
\ddot{{\mathbf{\hat{S}}}}^{\alpha }(\vec{q}
,t)+\sum_{\alpha ^{\prime }}
(\mbox{\boldmath {$\hat{\Omega}$}}^2(\vec{q}))^{\alpha
\alpha ^{\prime }}{\mathbf{\hat{S}}}^{\alpha ^{\prime }}(\vec{q},t)=0
\label{eqAs13}
\end{equation}
with the frequency matrix squared:
\begin{equation}
\mbox{\boldmath {$\hat{\Omega}$}}^2
(\vec{q})^{\alpha \alpha ^{\prime }} 
=
\left({\mathbf{J}}^{\alpha}(\vec{q})\right)^{\frac{1}{2}}
[{\mathbf{q}}^{\alpha }
{\mathbf{S}}^{-1}\left(\vec{q}\right) 
{\mathbf{q}}^{\alpha ^{\prime }}
-{\mathbf{C}}^{\alpha\alpha ^{\prime }}(\vec{q})]
\left({\mathbf{J}}^{\alpha ^{\prime}}(\vec{q})\right)^{\frac{1}{2}}. 
\label{eqAs14}
\end{equation}
Integration of (\ref{eqAs10}a) from $0$ to $t$ and substituting
$\int\limits_{0}^{t}dt^{\prime }{\mathbf{S}}^{\alpha'} (\vec{q},t')$ from
Eqs.~(\ref{eqAs10}b) with ${\mathbf{\hat{m}}}^{\alpha \alpha
^{\prime }}\equiv 0$ allows to express 
${\mathbf{\hat{S}}}(\vec{q},t)$ by $
{\mathbf{\hat{S}}}^{\alpha }(\vec{q},t)$:
\begin{eqnarray}
{\mathbf{\hat{S}}}(\vec{q},t)
& = &i{\mathbf{F}}(\vec{q}) 
{\mathbf{S}}^{-1}\left(\vec{q}\right) \sum_{\alpha
,\alpha ^{\prime }}{\mathbf{q}}^{\alpha }
\left[ \left( {\mathbf{C}}^{\beta \beta
^{\prime }}\left( \vec{q}\right) \right) ^{-1}\right] ^{\alpha \alpha
^{\prime }} 
\\\nonumber
& &
\times \left({\mathbf{J}}^{\alpha ^{\prime }}(\vec{q})\right) ^{1/2}
{\mathbf{\hat{S}}}^{\alpha ^{\prime }}(\vec{q},t)  \label{eqAs15}
\end{eqnarray}

Similar to 
\mbox{\boldmath {$\Omega $}}, the matrix 
$\left( \mbox{\boldmath {$\hat{\Omega }$}} ^{\alpha \alpha ^{\prime }}\right) 
$ is non-diagonal in $l,l^{\prime }$ and
in $\alpha $, $\alpha ^{\prime }$. This again leads to a coupling between
the translational and orientational modes, which are the eigenmodes of $
\left(\mbox{\boldmath {$\hat{\Omega }$}} 
^{\alpha \alpha ^{\prime }}\right) $. Now it is
(see \cite{schilling97}):
\begin{equation}
{\mathbf{C}}^{\alpha \alpha ^{\prime }}\left( \vec{q}\right) =\tilde{
{\mathbf{q}}}^{\alpha } 
\tilde{{\mathbf{C}}}^{\alpha \alpha ^{\prime }}\left( \vec{q}
\right) \tilde{{\mathbf{q}}}^{\alpha ^{\prime }}  \label{eqAs16}
\end{equation}
with:
\begin{equation}
\lim_{\vec{q}\rightarrow 0}\tilde{C}_{lm,l^{\prime }m^{\prime
}}^{\alpha \alpha ^{\prime }}\left( \vec{q}\right) \neq 0  \label{eqAs17}
\end{equation}
and
\begin{eqnarray}
\left(\tilde{{\mathbf q}}^{\alpha}\right)_{lm,l^{\prime }m'}
=
\tilde{q}^{\alpha}_{lml^{\prime }m'}
&=&\tilde{q}_{l}^{\alpha}
\delta _{ll^{\prime }}
\delta _{mm^{\prime }}
  \label{eqAs18} \\
\tilde{q_{l}}^{\alpha } &=&\left\{ 
\begin{array}{cc}
q, & \left( \alpha ,l\right) =\left( T,0\right)  \\ 
1, & \left( \alpha ,l\right) \neq \left( T,0\right),
\end{array}
\right.   
\end{eqnarray}
not to be confused with ${\mathbf{q}}^{\alpha }$. Therefore, we get from 
Eq.~(\ref{eqAs14}) with Eqs.~(\ref{eq6}), (\ref{eqAs4}), (\ref{eqAs16}) and
(\ref{eqAs18}):
\begin{eqnarray}
\label{eqAs19}
\left( {\mbox{\boldmath {$\hat{\Omega}$ }}}^{2}\left( \vec{q}\right) \right)
_{00,00}^{TT}=\frac{k_{B}T}{M}q^{2}\left( {\mathbf{S}}^{-1}\left( \vec{q}
\right) \right) _{00,00}
\\\nonumber
\left[ 1+\left( {\mathbf{S}}\left( \vec{q}\right) 
\tilde{{\mathbf{C}}}^{TT}\left( \vec{q}\right) \right) _{00,00}\right] 
\end{eqnarray}
for the acoustic part of $\left( ({\mbox{\boldmath{$\hat\Omega$}}}^{2})^{\alpha
\alpha ^{\prime }}\right)  $. Taking into account that we have not normalized 
$\mathbf{C}^{\alpha \alpha ^{\prime }}$ and ${\mathbf{\tilde{C}}}^{\alpha
\alpha ^{\prime }}$, the result Eq.~(\ref{eqAs19}) is completely analogous to
the result derived in \cite{goetze00}
for $\left( \hat{{\mathbf{\Omega }}}\left( 
\vec{q}\right) \right) ^{2}$ for simple one-component liquids. 
However, we note that in contrast to \cite{goetze00} the equations of
motion for the rescaled correlators ${\mathbf{\hat{S}}}(\vec{q},t)$ and $
{\mathbf{\hat{m}}}^{\alpha \alpha ^{\prime }}(\vec{q},t)$ are not covariant, due to
the splitting of the current density into a translational and rotational
part. Therefore $\left( \hat{\mbox{\boldmath{$\Omega$}}}^{2}\left( 
\vec{q}\right) \right)^{\alpha \alpha ^{\prime }}$ 
is not a straight forward generalization of 
$\hat{\mbox{\boldmath{$\Omega$}}}^{2}\left( \vec{q}\right) $ 
for simple liquids to molecular liquids.

The coupling between translational and orientational modes in the liquid and
in the glass already on the linear level of the equations of motion is not
surprising due to the interaction between translational and orientational
degrees of freedom. But, it is also obvious that memory effects will lead to
additional couplings and it is this point, which we will investigate in
section III. 

\subsection{Mode coupling theory}

\label{cap3.1}

In the preceding subsection we have neglected the memory kernels.
Approaching the glass transition significant memory effects occur. Therefore 
${\mathbf{m}}^{\alpha \alpha ^{\prime }}$must be taken into account. Using
mode-coupling theory (MCT) an approximate expression for the slow part $
{\mathbf{m}}^{\alpha \alpha ^{\prime }}(\vec{q},t)$ has been derived which
leads to a closed set of equations for ${\mathbf{S}}(\vec{q},t).$
This has been done for molecular systems using the molecular representation 
\cite{franosch97a,schilling97,fabbian99} and a site-site description
\cite{chong00}. For a liquid of
linear molecules we will use: 
\begin{eqnarray}
m_{lm,l^{\prime }m^{\prime }}^{\alpha \alpha ^{\prime }}(\vec{q}
,t) &\approx &\frac{\sqrt{I_{\alpha }I_{\alpha ^{\prime }}}}{k_{B}T}\Gamma
_{lm,l^{\prime }m^{\prime }}^{\alpha \alpha ^{\prime }}(\vec{q})\delta (t)+ 
  \label{eq35a} \\
&+&\left( m_{lm,l^{\prime }m^{\prime }}^{\alpha \alpha ^{\prime }}(\vec{q}
,t)\right) _{\text{slow}}
\end{eqnarray}
where the first and second term on the r.h.s. of Eq. (\ref{eq35a}) accounts
for the fast and the slow contributions, respectively. MCT yields \cite
{schilling97} 
\begin{eqnarray}
\left( m_{lm,l^{\prime }m^{\prime }}^{\alpha \alpha ^{\prime }}(\vec{q}
,t)\right) _{\text{slow}} 
=
\frac{1}{2N}
\sum_{\vec{q}_{1}\vec{q}_{2}\atop \vec{q}_{1}+\vec{q}_{2}=\vec{q}}
\sum_{l_{1}m_{1}\atop l_{2}m_{2}}
\sum_{l_{1}^{\prime }m_{1}^{\prime }\atop l_{2}^{\prime }m_{2}^{\prime }}
\times  \label{eq35b} \\
\times V_{lm,l^{\prime }m^{\prime }}^{\alpha \alpha ^{\prime }}(\vec{q}|
\vec{q}_{1}l_{1}m_{1},l_{1}^{\prime }m_{1}^{\prime };\vec{q}
_{2}l_{2}m_{2},l_{2}^{\prime }m_{2}^{\prime })   \\
\times  S_{l_{1}m_{1},l_{1}^{\prime }m_{1}^{\prime }}(\vec{q}
_{1},t)S_{l_{2}m_{2},l_{2}^{\prime }m_{2}^{\prime }}(\vec{q}_{2},t)  
\end{eqnarray}
The explicit expressions for the vertices ${\mathbf{V}}$ are given in Ref. 
\cite{schilling97}. Equations (\ref{eqAs2}) together with equations (\ref
{eq35a}) and (\ref{eq35b}) are a closed set of equations for $
S_{lm,l^{\prime }m^{\prime }}(\vec{q},t)$ which need the damping
coefficients $\Gamma _{lm,l^{\prime }m^{\prime }}^{\alpha \alpha ^{\prime }}(
\vec{q})$ and the static correlators $S_{lm,l^{\prime }m^{\prime }}(\vec{q})$
as an input. The latter uniquely determine the vertices.

It is obvious that the MCT-polynomial $\left( m_{lm,l^{\prime }m^{\prime
}}^{\alpha \alpha ^{\prime }}\right) _{\text{slow}}$ leads to additional
coupling between the correlators $S_{lm,l^{\prime }m^{\prime }}(\vec{q},t)$.
Due to this nonlinearity the MCT-equations (\ref{eqAs2}), (\ref{eq35a}) and (
\ref{eq35b}) can only be solved numerically. We have performed such a
numerical solution for dipolar hard spheres, one of the simplest systems
involving translational and rotational motion.

\subsection{Dipolar hard sphere liquid}

\label{cap3.2} The investigation of dipolar hard spheres (DHS) has the
advantage that approximate analytical expressions \cite{wertheim71} for the
static correlators are known.

We consider a system of N hard spheres with homogeneous number density $\rho$,
 diameter $d$
, mass $M$, moment of inertia $I=\frac{1}{10}Md^{2}$ and dipolar moment $\mu 
$. The origin of the body fixed frame is chosen to coincide with the center
of mass of each sphere, which is the natural choice. Since the density of
the particles is homogeneous, the center of mass is equal to the center of
the spheres. 
The reader should note that the MCT-equations (\ref{eqAs2}), (\ref{eq35a}) and  (\ref{eq35b})
and even the original exact equations of motion Eqs. (\ref{eqAs2}) 
are not covariant (i.e. invariant in their form) under a shift
of the reference point for the body-fixed frame. In order to get covariant
equations one has to project on the individual Cartesian components of
$\vec{j}^T_{lm}(\vec{q})$ and $\vec{j}^R_{lm}(\vec{q})$. Nevertheless, we
think that the present equations are a reasonable good approach to the
dynamics of molecular liquids, because of the natural choice of the reference
frame.  
Of course, this point needs additional
investigations.
The advantage of MCT in site-site representation \cite{chong00} is
that this problem does not occur, because no reference point must be
chosen.

The physical control parameters are the packing fraction $\phi =\frac{\pi }{6
}\rho d^{3}$ and the temperature $T$. In
the following the length unit is chosen such that $d=1$. In addition we
choose $M=1$ and $\mu =1$. This choice means that time $t$ and temperature $T
$ are measured in units of $M^{1/2}d^{5/2}/\mu $ and $k_{B}^{-1}\mu ^{2}/d^{3}
$, respectively. In the following we will use $T^{\ast }=T/(\mu
^{2}/k_{B}d^{3})$ as dimensionless temperature.

As already stated, the MCT-equations (\ref{eqAs2}), (\ref{eq35a}) and (\ref
{eq35b}) require $\Gamma _{lm,l^{\prime }m^{\prime }}^{\alpha \alpha
^{\prime }}(\vec{q})$ and $S_{lm,l^{\prime }m^{\prime }}(\vec{q})$ as an
input. Throughout this paper we will put all damping coefficients $\Gamma
_{lm,l^{\prime }m^{\prime }}^{\alpha \alpha ^{\prime }}$ to zero. The static
correlators $S_{lm,l^{\prime }m^{\prime }}(\vec{q})$ are obtained from
Wertheim's solution who used Percus-Yevick and mean spherical approximation 
\cite{wertheim71} This leads to 
\begin{equation}
S_{lm,l^{\prime }m^{\prime }}(\vec{q})\approx \left\{ 
\begin{array}{l@{\quad,\quad}l}
S_{lm}(q)\delta _{ll^{\prime }}\delta _{mm^{\prime }} & l=0,1\,,\,l^{\prime
}=0,1 \\ 
\delta _{ll^{\prime }}\delta _{mm^{\prime }} & \text{otherwise}
\end{array}
\right.   \label{eq36}
\end{equation}
The vertices of the memory kernels are bilinear in the direct correlation
functions 
\begin{equation}
c_{lm}(q)=\frac{4\pi }{\rho }\left[ 1-\frac{1}{S_{lm}(q)}\right] 
\label{eq38}
\end{equation}

Since we have chosen the so-called $q$-frame \cite{gray84} (for which $
\vec{q}=$ $\vec{q}_{0}\equiv \left(
0,0,q\right) $) the correlators and the kernels 
become diagonal in $m$ and $m^{\prime }$ 
\cite{schilling97}. The fact that the static correlators are diagonal in 
$l$ and $l^{\prime }$ and are structureless for $l$, $l^{\prime }\geq 2$ is
an artefact of both approximations. Additional shortcomings are (i) the
independence of the center of mass correlator $S_{00}\left( q\right) $ on
temperature and (ii) the smooth behavior of $S_{lm}\left( q\right) $ at $
\phi _{\text{rcp}}\simeq 0.64$, the value for random close packing. The
static correlators from Wertheim's approach exhibit a divergency at $\phi
=\phi _{\max }=1$, only. We will come back to these points below.

Due to (\ref{eq36}) and (\ref{eq38}) only mode coupling occurs between the
time dependent correlators with $l_{1},l_{1}^{\prime },l_{2},l_{2}^{\prime }$
smaller or equal to one.
In order to simplify the
MCT-equations as much as possible we use the additional approximations (in
the $q$-frame) 
\begin{equation}
S_{lm,l^{\prime }m^{\prime }}(\vec{q},t)\approx \left\{ 
\begin{array}{l@{\quad,\quad}l}
S_{lm}(q,t)\delta _{ll^{\prime }}\delta _{mm^{\prime }} & 
l=0,1\,,\,l^{\prime }=0,1 \\ 
0 & \text{otherwise}
\end{array}
\right.  \label{eq39a}
\end{equation}

\begin{equation}
m_{lm,l^{\prime }m^{\prime }}^{\alpha \alpha ^{\prime }}(\vec{q},t)\approx
\left\{ 
\begin{array}{l@{\quad,\quad}l}
m_{lm}^{\alpha \alpha ^{\prime }}(q,t)\delta _{ll^{\prime }}\delta
_{mm^{\prime }} & l=0,1\,,\,l^{\prime }=0,1 \\ 
0 & \text{otherwise}
\end{array}
\right.   \label{eq39b}
\end{equation}
In section IIB we have shown that translational and orientational dynamics
is already coupled on the linear level due to non-diagonality in $l$ and $
l^{\prime }$. Above we have chosen all correlators to be diagonal in $l$ and 
$l^{\prime }$. Therefore, a coupling between $S_{00}\left( q,t\right) $ and $
S_{lm}(q,t)$ with $l>0$ can only originate from mode-coupling effects.
Therefore, our diagonalization approximation (which is even exact in
our case due to the restriction of $l$ and $l'$ to $0$ and $1$) allows
to study the influence
of the mode-coupling terms on the microscopic dynamics without interfering
with the direct coupling mechanism between the correlators $S_{lm}(q,t)$
discussed in section IIB.

The memory kernels (cf. Eq. (\ref{eq35b})) in the $q$-frame $m_{lm,l^{\prime
}m^{\prime }}^{\alpha \alpha ^{\prime }}(\vec{q},t)=m_{lm}^{\alpha \alpha
^{\prime }}(q,t)\delta _{ll^{\prime }}\delta _{mm^{\prime }}$ contain the
following couplings \cite{schilling97}: 
\begin{equation}
m_{00}^{\alpha \alpha ^{\prime }}(q,t)\leftrightarrow
S_{00}(q_{1},t)S_{00}(q_{2},t)\text{ and }
S_{1m_{1}}(q_{1},t)S_{1m_{2}}(q_{2},t)  \label{eq37}
\end{equation}
\begin{equation}
m_{1m}^{\alpha \alpha ^{\prime }}(q,t)\leftrightarrow
S_{00}(q_{1},t)S_{1m_{2}}(q_{2},t)\text{ and }
S_{1m_{1}}(q_{1},t)S_{00}(q_{2},t)  \label{eq37a}
\end{equation}
Eq. (\ref{eq37}) shows that the center of mass correlator $S_{00}(q,t)$ may
undergo a glass transition independently from the dipoles whereas Eq. (\ref
{eq37a}) demonstrates that the dipoles are ''slaved'' by the center of mass
dynamics and can freeze only if $S_{00}(q,t)$ has become non-ergodic. In
order that $S_{00}(q,t)$ and $S_{1m}(q,t)$ freeze simultaneously the
vertices of the bilinear terms in Eq. (\ref{eq37a}) must be large enough.
This happens at low enough temperatures. Since $S_{l,-m}(q,t)\equiv
S_{l,m}(q,t)$ \cite{schilling97}, there are three independent correlators
which are the center of mass correlation function $S_{00}(q,t)$ and two
dipolar ones $S_{1m}(q,t)$, $m=0,1$. In the following we consider the
normalized correlators $\phi _{lm}(q,t)=S_{lm}(q,t)/S_{lm}(q)$.

\begin{figure}
	\centerline{\rotatebox{-0}{\resizebox{9cm}{!}
	{\epsfig{file=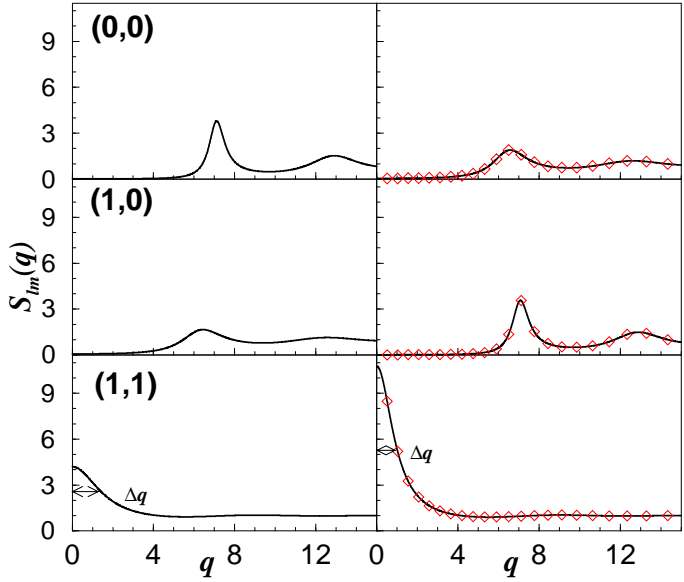,width=8cm,angle=-0}}}}
	\caption{The $q$ dependence of the static structure factors
	$S_{lm}(q)$ for $(lm)=(0,0),\, (1,0),\, (1,1)$ and two different pairs
	of $(\phi,T^*)$ indicated by the full circles in Figure \ref{fig3}:
	 $(0.53,0.3)$ (left), $(0.381,0.04)$ (right).
	$\Delta q$ is the peak width of $S_{11}(q)$ at half maximum.
	The symbols indicate the discretized $q$-values (see text)}
	\label{fig1}
\end{figure}

The three static correlators are shown in Fig.~\ref{fig1} for two different
pairs of $\left( \phi ,T^*\right) $. Three main features can be seen. First,
the $q$-variation of $S_{00}$ of the ``longitudinal '' dipolar correlator
and $S_{10}$ is rather similar to each other and resembles that of a simple
liquid with a well pronounced main peak at $q_{\max }\approx 2\pi /\overline{
a}$, where $\overline{a}\approx d=1$, the mean distance between nearest
neighbors. The ``transversal'' dipolar correlator $S_{11}\left( q\right) $
behaves quite differently. It exhibits only one peak which is at $q=0$ and
it becomes almost structureless for $q>\triangle q$, the peak width at
half maximum.  Second, $S_{11}\left( q=0\right) $ increases with decreasing
temperature, whereas $\triangle q$ decreases. This behavior signals a
medium range orientational order due to a precursor of ferrofluid order. This
order is induced by an enhancement of the dipolar interactions at lower
temperatures with respect to the hard core repulsion. Third, $S_{1m}\left(
q=0\right) $ depends on $m$, because of the long range nature of the dipolar
interactions.

Besides the truncation at $l=1$ we must also truncate and discretize the $q$
-variable. Since the reduction (for $N\rightarrow \infty $) to a \emph{single
} integral of the sum over $\vec{q}_{1}$ and $\vec{q}_{2}$ in Eq. (\ref
{eq35b}) for simple liquids \cite{latzphd} cannot easily be used for
molecular liquids the number of steps to calculate this sum increases
quadratically with the number of $q$-values, instead of a linear increase
for simple liquids. This fact makes the numerical solution of the molecular
MCT-equations rather CPU-time consuming. Therefore we decided to choose a
non-equidistant distribution of 30 $q$-values between $q_{low}\simeq 0.51$
and $q_{up}\simeq 40$. These values were generated by the non-linear
relation 
\begin{equation}
q_{\nu }=\frac{1}{\alpha }\text{arctanh}(\nu \Delta _{x}),\quad \nu
=1,2,...,30.  \label{eq40}
\end{equation}
with $\Delta _{x}=\frac{\text{tanh}{(\alpha q_{co})}}{31}$ and $q_{co}=50$.
The parameter $\alpha $ has been chosen such that the main peak of, e.g. $
S_{00}(q)$, is still in the linear regime of arctanh. The right of Fig. \ref
{fig1} demonstrates this for $\alpha =0.065$. It is
obvious from (\ref{eq40}) that for $q<\alpha ^{-1}$ the $q_{\nu }$ are
almost equidistant and for $q>\alpha ^{-1}$ they become more and more
diluted. Since the large $q$-regime is not as important as the range around
the main peak in $S_{00}(q)$ or that at $q=0$ in $S_{11}(q)$ (which drive
the glass transition) our choice for $q_{\nu }$ should not influence our
results, at least qualitatively. We have solved the molecular MCT-equations (
\ref{eqAs2}), (\ref{eq35a}) and  (\ref{eq35b}) 
in time space using an algorithm already developed to solve the
MCT-equations for simple liquids \cite{fuchs91b}. However, our numerical
procedure differs from that by G\"{o}tze et al. \cite{goetze00c} for the single
dumbbell in an isotropic liquid. 
These authors introduced an effective memory kernel (independent on $\alpha $
and $\alpha ^{\prime }$) and an effective microscopic frequency.

\begin{figure}	
\centerline{\rotatebox{-0}{\resizebox{9cm}{!}
	{\includegraphics{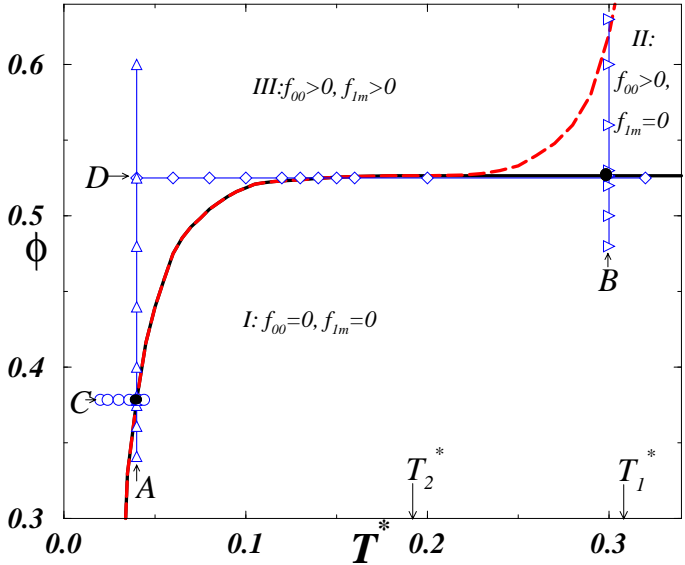}}}}
	\caption{Glass transition phase diagram for dipolar hard spheres:
	 The solid and the dashed line are critical lines
	at which a 
	discontinuous (type B) and a continuous (type A) glass transition
	takes place between the phases I,II and III. 
	A, B, C and D denote the various paths on which we have
	investigated the control parameter dependence of the dynamics. 
	The full circles indicate the two points at which the static
	correlators in Figure \ref{fig1} and the microscopic
	frequencies in Figure \ref{fig4} were calculated.
	For $T_1^*$ and $T_2^*$ see text}
\label{fig3}
\end{figure}

Before we come to
the dynamics let us shortly discuss the phase diagram for a glass transition
which was already calculated \cite{schilling97,scheidsteger98}. 
Throughout the rest
of this paper all the results are given in the $q$-frame.
The modification with respect to Refs. \cite{schilling97,scheidsteger98} of the $
q$-discretization leads to a small 
quantitative change of the glass transition
lines, but without changing the topology. Therefore we have calculated again
non-ergodicity parameters 
\begin{equation}
f_{lm}(q)=\lim_{t\rightarrow \infty }\phi _{lm}(q,t)  \label{eq41}
\end{equation}
as a function of $\phi $ and $T^{\ast }$ from which the phase diagram is
obtained (Figure \ref{fig3}). Since $\phi $ can not exceed $\phi
_{rcp}\simeq 0.64$, the value for random close packing of hard spheres, we
have plotted only $\phi \leq \phi _{rcp}$. There are two significant
temperatures $T_{1}^{\ast }$ and $T_{2}^{\ast }$. At $T_{1}^{\ast }$ the
critical line $\phi _{\text{type A}}^{c}(T^{\ast })$ (dashed line in Figure 
\ref{fig3}) reaches $\phi =\phi _{rpc}$ and at $T_{2}^{\ast }$ it merges
into the critical line $\phi _{\text{type B}}^{c}(T^{\ast })$ (solid line in
Figure \ref{fig3}). These two temperatures have the following meaning. For $
T^{\ast }>T_{1}^{\ast }$ an increase of $\phi $ leads at $\phi _{\text{type B
}}^{c}(T^{\ast })$ to a glass transition for the center of mass motion, but
not for the dipoles. Choosing $T^{\ast }$ between $T_{1}^{\ast }$ and $
T_{2}^{\ast }$ again a glass transition of the center of mass motion occurs
at $\phi _{\text{type B}}^{c}(T^{\ast })$. But increasing $\phi $ beyond $
\phi _{\text{type B}}^{c}(T^{\ast })$ a spin glass like transition for the
dipoles occurs at $\phi _{\text{type A}}^{c}(T^{\ast })$. Below $T_{2}^{\ast
}$, center of mass and dipolar dynamics freeze simultaneously at $\phi _{
\text{type B}}^{c}(T^{\ast })$. For $T^{\ast }>T_{2}^{\ast }$ the critical
value $\phi _{\text{type B}}^{c}(T^{\ast })$ is identical to that for hard
spheres $\phi _{HS}^{c}$. For the present choice of $q$-discretization we
find $\phi _{HS}^{c}\simeq 0.5265$ which does not differ much from the more
precise value $\phi _{HS}^{c}\simeq 0.516$ obtained with $100$ equidistant $q
$-values \cite{franosch97}.

The glass transition at high and low temperatures is driven by different
physical mechanisms. At high temperatures it is the so-called cage effect,
which leads to the freezing of the liquid into a non-ergodic
phase, due to an
increase with increasing density of the main peak of the center of mass
correlator $S_{00}\left( q\right) $. Lowering the temperature enhances the
role of the dipolar interactions which leads to a strong increase of the
peak at $q=0$ of the ``transversal'' dipolar correlator $S_{11}\left(
q\right) $. The increase of the dipolar correlations are accompanied by a
decrease of the center of mass correlations (cf.~Fig.~\ref{fig1}), and take
over the role of the cage effect. This behavior reflects itself in the
non-ergodicity parameters. $f_{11}\left(
q=0\right) $ increases with decreasing temperature much stronger than $
f_{00}\left( q_{\max }\right) $. 
For details the reader is referred to \cite{schilling97}.

The properties of the phase diagram require some more comments: (i)
the $T^*$-independence of $\phi _{\text{type B}}^{c}(T^*)$ for $T^*>T_{1}^{\ast }$
originates from the $T^*$-independence of $S_{00}\left( q\right) $, which is
an artefact of Wertheim's solution. 
Removing this artefact will result in a $
T^*$-dependence of $\phi _{\text{type B}}^{c}(T)$ for {\em all} $T^*$ with 
$\lim_{T^*\rightarrow \infty }\phi _{\text{type B}}^{c}(T^*)=\phi
_{HS}^{c}$, 
since dipolar interactions will be irrelevant at infinite
temperature 
(ii), 
the existence of $T_{1}^{\ast }$ relies on the fact that $
\phi _{\text{type A}}^{c}(T^*)$ approaches the value $\phi _{\text{rcp}}$ at a 
\textit{finite} temperature. This is true when the static correlators within
Percus-Yevick approximation are used. Whether or not an improved theory
reproducing the singular behavior at $\phi _{\text{rcp}}$ would lead to the
same conclusion is not obvious. Since such a theory does not exist, we can
not exclude that $\phi _{\text{type A}}^{c}(T^*)$ only approaches $\phi _{
\text{rcp}}$ at $T^*=\infty $.

Now we turn to the time or frequency dependent features of dipolar hard
spheres. We have studied the control parameter dependence of the dynamics 
along the paths A, B, C and D,
indicated in Figure \ref{fig3}. In contrast to the determination of the long
time behavior inertia effects will play an important role in the microscopic
time or frequency regime. In the liquid, these inertia effects enter through
the microscopic frequency matrix $\mbox{\boldmath {$\Omega
$}}(\vec{q})$. Due to the diagonality of ${\mathbf{S}}(\vec{q})$ the matrix $
\mbox{\boldmath {$\Omega
$}}(\vec{q})$ becomes diagonal with diagonal elements (in $q$-frame) which
follows from Eq. (\ref{eq17b}): 
\begin{equation}
\Omega _{lm}(q)=\sqrt{\left[ \frac{k_{B}T}{M}q^{2}+\frac{k_{B}T}{I}l(l+1)
\right] /S_{lm}(q)}\quad ,  \label{eq42}
\end{equation}
restricted to $l=0,1$. Figure \ref{fig4} depicts $\Omega _{lm}(q)$ for two
different pairs $(\phi ,T^{\ast })$ in vicinity of the critical line $\phi _{
\text{type B}}^{c}(T^{\ast })$. This Figure reveals two features which stem
from the properties of $S_{lm}(q)$ discussed above. First, $\Omega _{1m}(q=0)
$ depends on $m$. Second, and most important, a ''softening'' of the
''optic'' frequency $\Omega _{11}(q)$ with $q\rightarrow 0$ is observed for
decreasing temperature $T^{\ast }$ for which the role of dipolar
interactions (compared to the hard core repulsion) becomes more and more
enhanced. This ``softening'' comes from the strong increase of $S_{11}(q)$
for $q\rightarrow 0$ (see Fig. \ref{fig1})\ and has its physical origin in the
occurrence of a \emph{medium range orientational order} of the dipoles. The
inverse width $(\Delta q)^{-1}$ of the peak in $S_{11}(q)$ at $q = 0
$ is a measure for the length scale for this orientational order. $(\Delta
q)^{-1}$ increases with decreasing temperature (see Fig. \ref{fig1}). We
will come back to this point in the final section.

The microscopic frequencies in the glass phase are given by the
``renormalized'' frequency matrix $\left( \left( \hat{\mbox{\boldmath{$\Omega$ }}}
^{2}\left( \vec{q}\right) \right) ^{\alpha \alpha ^{\prime
}}\right) $. Taking again into account the diagonalization with respect to $l
$ and $l^{\prime }$ we obtain from Eq.~(\ref{eqAs14}) with Eqs.~(\ref{eqAs4})
and (\ref{eqAs16})
\begin{eqnarray}
\left( \hat{\mathbf{\Omega }}_{lm}^{2}\left( q\right) \right)
^{\alpha\alpha ^{\prime }}
& = &
\frac{k_{B}T}{\left( I_{\alpha }I_{\alpha }^{\prime
}\right) ^{1/2}}\frac{1}{S_{lm}\left( q\right) }\times
\\\nonumber
& \times & \left[ q_{l}^{\alpha
}q_{l}^{\alpha ^{\prime }}+\tilde{q}_{l}^{\alpha }\tilde{q}
_{l}^{\alpha ^{\prime }}\tilde{C}_{lm}^{\alpha \alpha ^{\prime }}\left(
q\right) S_{lm}\left( q\right) \right] \quad .  \label{eqAs20}
\end{eqnarray}

\bigskip Since $\tilde{C}_{00}^{\alpha \alpha ^{\prime }}\left( q\right) 
$ is different from zero for $\alpha =\alpha ^{\prime }=T$, only, we get with
Eqs.~(\ref{eq6}), (\ref{eqAs8}),  (\ref{eqAs16}) and (\ref{eqAs18}
):
\begin{equation}
\hat{\mathbf{\Omega }}_{00}\left( q\right) =\Omega _{00}\left(
q\right) /\left( 1-f_{00}\left( q\right) \right) ^{1/2}\quad .  \label{eqAs21}
\end{equation}

$\tilde{C}_{1m}^{\alpha \alpha ^{\prime }}\left( q\right) $ is nonzero
for all $\left( \alpha \text{, }\alpha ^{\prime }\right) $. Therefore two
eigenfrequencies $\hat{\Omega}_{1m}^{\pm }\left( q\right) $ exist for each $m$. $
\hat{\mathbf{\Omega }}_{00}\left( q\right) $ and $\hat{\Omega} _{1m}^{\pm
}\left( q\right) $ are shown in Fig.~\ref{fig4a}.
the ''renormalized'' frequencies require the non-ergodicity parameters as an
input. The latter become less accurate for small $q$, because the
discretization of the $q$-values influences the results much stronger for
decreasing $q$. Therefore we do not present data below $q=1$. 
Of course it is $\hat{\Omega}_{00}(q) \to \hat{c}q$ and $\hat{\Omega}^\pm_{1m}(q)
\to  \hat{\Omega}^\pm_{1m}(0)>0$ for $q\to 0$. Note that (i)
$\hat{\Omega}^\pm_{11}(q)$ does not vary much with $q$ and (ii) that
$\hat{\Omega}^-_{11}(q)\ll\hat{\Omega}_{00}(q_{max}\simeq 7)$.  

Now let us turn to the equations (\ref{eqAs2}) including the time-dependent
memory kernels given by  (\ref{eq35a}) and (\ref{eq35b}). 
From the solutions of these MCT-equations we get the
normalized correlators $\phi _{lm}(q,t)$ or equivalently the corresponding
susceptibility spectra $\chi _{lm}^{\prime \prime }(q,\omega )$ related to
the correlation spectra $\phi
_{lm}^{\prime \prime }(q,\omega )$ 
by $\chi _{lm}^{\prime \prime }(q,\omega ) =\omega \phi
_{lm}^{\prime \prime }(q,\omega )$. Figures \ref{fig5} and \ref{fig6} show
$\phi_{lm}(q,t)$ and $\chi_{lm}''(q,\omega)$ 
for $q\simeq 4.7$ along path A in Figure \ref{fig3}. Figure \ref
{fig5} clearly demonstrates for all correlators the formation of a plateau
and the tremendous slowing down of the relaxation with increasing $\phi$
which stops at $\phi_{c}(T^*)\simeq 0.3782$. 

The dynamical
features are better recognized in $\omega $-space.
Before we discuss these results, let us comment on the quality of our $
\omega $-dependent data. Because the glassy dynamics involves many decades
in time one has to use a special algorithm, which decimates the time step
with increasing time, i.e.~one chooses an initial time increment $h$ to
discretize the time axis \cite{fuchs91b}. After a certain number of time steps the
increment is doubled, etc. Although the $t-$dependent data shown in Fig.~\ref
{fig5} look continuous, they exhibit tiny discontinuities of the
correlators and of their first time derivative at those times at which the
increment is doubled. These discontinuities which could be diminished by an
increase of the CPU-time lead to the wiggles which can,
e.g.~be seen in Fig.~\ref{fig6}. However, we have checked by variation of
the parameters $h$, the boundary values of the Fourier integrals, etc. that the
features we will address below are not influenced by the wiggles. 
Further,
the low frequency wing of the $\alpha -$peak for $\chi _{lm}^{\prime \prime
}(q,\omega )$ in Fig.~\ref{fig6} shows a rather small deviation from 
a linear $\omega$-dependence, as it should be. 
This failure is due to the choice of the
lower bound for the Fourier integral. If this is taken to be zero, the
wiggles become more pronounced. Therefore, choosing a nonzero bound is a
compromise between the reduction of the wiggles and a small 
deviation from the
linearity in $\omega $ for $\omega \rightarrow 0$. We have also used a
spline-technique for smoothing the data in time space. The Fourier
transform of those has reproduced all the relevant features obtained from
the original $t-$dependent correlators.

Let us now discuss the spectra.
For $\phi<\phi_c(T_A^*)$, i.e. in the {\em liquid phase}, Figure \ref{fig6}
left column
reveals the existence of three peaks for $\chi_{00}''(q\simeq 4.7,\omega)$
and  
two peaks for $\chi_{11}''(q\simeq 4.7,\omega)$ and
$\chi_{10}''(q\simeq 4.7,\omega)$. The low frequency peak in all
spectra $\chi_{lm}''(q\simeq 4.7,\omega)$
is the $\alpha$-peak related to the slowing down of the structural relaxation
(translational and rotational degrees of freedom). 
$\chi_{00}''(q\simeq 4.7,\omega)$ has a
high-frequency (hf) peak at $\omega_{hf}\approx 10$ and an additional peak at
$\omega_{op}\simeq 1$, i.e. about one decade below the hf-peak. 
$\chi_{10}''(q\simeq 4.7,\omega)$  and $\chi_{11}''(q\simeq 4.7,\omega)$ 
exhibits only one peak in the microscopic
frequency domain at $\omega\approx \omega_{op}$. 
Below we will show that the peak in $\chi_{00}''(q\simeq 4.7,\omega)$ at
$\omega\approx \omega_{op}$ originates from the orientational dynamics.
Therefore we call it ''orientational'' peak (op).

The situation for $\phi>\phi_c(T_A^*)$, i.e. in the {\em glass phase}, is
similar with two exceptions (cf. right column of Fig. \ref{fig6}). 
First, it is clear that there is no $\alpha$-peak anymore and second an
additional peak  located between the op and the hf-peak exists in 
$\chi_{00}''(q\simeq 4.7,\omega)$ for
$\phi=0.53$ and $0.60$, i.e. deep in the glass phase.
Even a further "peak" looking more like a shoulder appears for
$\phi=0.6$. Such additional peaks can also be seen in 
$\chi_{10}''(q\simeq 4.7,\omega)$  and $\chi_{11}''(q\simeq 4.7,\omega)$
(cf. right column of Fig. \ref{fig6})

In the
following we will concentrate ourselves onto the microscopic regime.
Therefore we can use a linear $\omega -$scale, which allows us to recognize
the $\omega$- and $\phi$-dependence (which we have described above) much
better for $0.1<\omega<10$. 
This is done in Figure \ref{fig7n} for the data from Figure \ref{fig6}. 
Both microscopic peaks at $\omega_{op}\approx 1$ and $\omega_{hf}\approx 10$
can clearly be seen in the compressibility spectrum
$\chi_{00}''(q\simeq 4.7,\omega)$. 
Whereas the intensity of the hf-peak increases with increasing $\phi$, the op
becomes more pronounced when approaching the glass transition from the liquid
side.
In the glass phase at $\phi=0.44$ and $\phi=0.48$ it is less prominent but
again becomes more pronounced for $\phi>0.5$, although its intensity
decreases. 
The appearance of the intermediate peak for $\phi=0.53$ and $\phi=0.60$ at
$\omega\approx6$ and $\omega\approx8$, respectively, can be observed
as well as the shoulder at $\omega\approx 4$.
Due to the linear $\omega$-scale we clearly 
see that  $\chi_{10}''(q\simeq 4.7,\omega)$
also exhibits an intermediate peak  for $\phi=0.53$ and $\phi=0.60$,
i. e. $\chi_{10}''(q\simeq 4.7,\omega)$ resembles
$\chi_{00}''(q\simeq 4.7,\omega)$ but with the opposite $\phi$-dependence of the
intensity of the op and hf-peak for $\phi>\phi_c(T^*_A)$.
$\chi_{11}''(q\simeq 4.7,\omega)$ which is proportional to the dielectric loss
$\epsilon''(q\simeq 4.7,\omega)$, possesses only one well pronounced peak at
$\omega\approx\omega_{op}$. 
To check how far the features depend on the path through a critical point, we
have investigated $\chi_{lm}''(q\simeq 4.7,\omega)$ along path C (see Figure
\ref{fig1}) for which $\phi=\phi_c=0.3786$ is fixed.
This allows to study how far density and temperature variation lead to similar
or different susceptibility spectra.
The results for path C are given in Figure \ref{fig8n}. 
$\chi_{00}''(q\simeq 4.7,\omega)$ shows an op and a hf-peak at
$\omega_{op}\approx 1$ and $\omega_{hf}\approx 10$, respectively. Their
corresponding intensities behave qualitatively similar for decreasing $T^*$ as
we have found along path A for increasing $\phi$. Intermediate peaks between
op and hf-peak do not occur.
$\chi_{10}''(q\simeq 4.7,\omega)$ differs from the result along path A,
whereas $\chi_{11}''(q\simeq 4.7,\omega)$ looks similar.

Translational and orientational dynamics freeze on path A and C
simultaneously.
Therefore we have also calculated $\chi_{lm}''(q\simeq 4.7,\omega)$ on path B
for fixed $T^*=0.30$ where $\phi_{00}(q,t)$ freezes at $\phi^{HS}_c \simeq
0.5265$ first, whereas $\phi_{1m}(q,t)$ undergoes a spin-glass-like transition
at $\phi_{\text{type A}}^c(T^*=0.3)\simeq 0.62$. 
The corresponding results are presented in Figure
\ref{fig9n}. $\chi_{00}''(q\simeq 4.7,\omega)$  and
$\chi_{1m}''(q\simeq 4.7,\omega)$ exhibit a well pronounced main peak. Its
position depends sensitively on $\phi$ in case of
$\chi_{00}''(q\simeq 4.7,\omega)$ and is practically $\phi$ independent for
$\chi_{1m}''(q\simeq 4.7,\omega)$. The latter quantity also pocesses a small
peak at $\omega\approx 10$, which originates from the translational motion via
translation-rotation-coupling. 

A peak in the susceptibility spectra can be either of relaxational or
oscillatory origin.
To distinguish between both types of behavior one must study the
$\omega$-dependence of the correlation spectra $\phi_{lm}''(q,\omega)$.
Spectra from neutron scattering are superpositions of $\phi_{lm}''(q,\omega)$
\cite{theis99}. 
A peak in $\phi_{lm}''(q,\omega)$ at $\omega>0$ proves the existence of an
oscillation whereas a peak at $\omega=0$ is of purely relaxational type.
The corresponding peak width is a measure of the damping. 
These correlation spectra are shown in Figure \ref{fig10n} and \ref{fig11n}
for path A and path B, respectively. 
For $\phi<\phi_c(T^*)$ (not shown in Fig. \ref{fig10n} and \ref{fig11n}) 
there is {\em no} peak at nonzero frequency, except for the hf-peak.
Therefore the op we have found in $\chi_{lm}''(q\simeq 4.7,\omega)$ in the
liquid phase is a purely relaxational excitation.
Now let us discuss the correlation spectra in the glass phase.
Figure \ref{fig10n} shows that in the glass, but close to the critical packing
fraction, there is a hf-peak at $\hat{\omega}_{hf}\approx 10$.
Deeper in the glass the position of that hf-peak shifts to higher
frequencies and an op at $\hat{\omega}_{op}\approx 1$ appears. Decreasing $\phi$
even more an intermediate peak between the op and the hf-peak is produced as well.
The position of the op and intermediate peak shifts to higher frequencies
with decreasing $\phi$ similar to the hf-peak, due to an increase of the
glass-stiffness.
The $\phi$-dependence of $\phi_{lm}''(q\simeq 4.7,\omega)$ at much higher
temperature $T^*=0.3$ (path B in Figure \ref{fig1}) is presented in Figure
\ref{fig11n}. 
We observe a similar behavior as in Figure \ref{fig10n}, i.e. besides the
hf-peak at $\hat{\omega}_{hf}=20-60 $ for $0.56\leq \phi\leq0.63$ there occurs
the op at $\hat{\omega}_{op}\approx 1$ and for $\phi=0.60$ and $\phi=0.63$ a
shoulder at about $\omega\approx 20$ which corresponds to the intermediate
peak in Figure
\ref{fig10n}. 

One of the conclusions we can draw from $\chi_{lm}''(q\simeq 4.7,\omega)$ and 
 $\phi_{lm}''(q\simeq 4.7,\omega)$ is that there is an op at $\omega_{op}\approx
 \hat{\omega}_{op}\approx 1$ roughly one decade below a hf-peak.
The hf-peak relates to a damped oscillation for the investigated range of
 $T^*$ and $\phi$, whereas the op changes its character from purely
 relaxational to damped oscillational behavior under going from the liquid to
 the glass.
Since the dipolar spectra  $\chi_{1m}''(q\simeq 4.7,\omega)$ and
 $\phi_{1m}''(q\simeq 4.7,\omega)$ have a main peak at about $\omega\approx 1$
 it is tempting to associate the op in  $\chi_{00}''(q\simeq 4.7,\omega)$  and
  $\phi_{00}''(q\simeq 4.7,\omega)$ with the orientational degrees of freedom
 and their coupling (via mode coupling effects) to the translational ones.  
If this interpretation is correct then
the op must exhibit an isotope effect with respect to a change of the moment
of inertia $I$. 
Figure \ref{fig12n} presents $\chi _{lm}^{\prime \prime
}(q\simeq 4.7,\omega )$ at $T_{A}^{\ast }=0.04$, $\phi
=0.381$ for $I=\frac{1}{10}$ (the correct value for a hard sphere with M=1
and d=1 and homogeneous mass distribution), $1$ and $10$. From this result
we find 
\begin{equation}
\label{eq42a}
\omega _{op} \propto \frac{1}{\sqrt{I}}.
\end{equation}
This type of isotope effect also occurs for $\phi _{lm}^{\prime \prime
}(q,\omega )$ which is shown in Figure \ref{fig13n} for $\phi _{lm}^{\prime
\prime }(q\simeq 4.7,\omega )$ at $T_{A}^{\ast }=0.04$, $\phi =0.530$ and for $I=
\frac{1}{10},1$ and $10$. This result yields: 
\begin{equation}
\label{eq42b}
\hat{\omega} _{op}\propto \frac{1}{\sqrt{I}}.
\end{equation}
The approximate scaling of $\omega _{op}$ and $\hat{\omega}_{op}$ with $
1/\sqrt{I}$ strongly supports the orientational origin of the op. Since $
\omega _{hf}$ and $\hat{\omega}_{hf}$ are rather insensitive on a change of $I$
their origin must lie in the translational motion.

So far we have not studied the $q$-dependence of all of these peaks. Since
this may elucidate more the features of the several peaks, we present $
\chi_{lm}^{\prime\prime}(q,\omega)$ at $T^*_A=0.04$, $\phi=0.381$ in the
glass but near the glass transition and $\phi_{lm}^{\prime\prime}(q,\omega)$
at $T^*_A=0.04$, $\phi=0.53$ deeper in the glass in Figures \ref{fig14n} and 
\ref{fig15n}, respectively. The result for $
\chi_{00}^{\prime\prime}(q,\omega)$ shows a nearly $q$-independent position
of the op at $\omega_{op}\approx 1$ for $1.0 \le q\le 10.6$, whereas the
position $\omega_{hf}$ of the hf-peak changes with $q$. A similar $q$
-sensitivity holds for the position of the main microscopic peak in $
\chi_{1m}^{\prime\prime}(q,\omega)$ for $m=0$ and $1$. The result for $
\phi_{lm}^{\prime\prime}(q,\omega)$ yields the same $q$-independence of the
op at $\hat{\omega}_{op}\simeq 1$ 
and a high sensitivity of the position of the hf-peak at $
\hat{\omega}_{hf}\approx 10$. The high frequency peak is practically
absent within a ''window'' around $q=6.5$ and it appears below and above that
window.

Finally, let us comment on the dependence of the peaks on $T^*$ and $\phi$
as well as on $\alpha$, which characterizes the distribution of the values
for $q$ (cf. Eq. (\ref{eq40}) and Figure \ref{fig1}). Let us start with the $
\alpha$-dependence. Figure \ref{fig16n} presents $\chi_{00}^{\prime\prime}(
\bar{q}_{max},\omega)$ at $T_A^*=0.04$, $\phi=0.381$ for three different
values of $\alpha $. $\bar{q}_{max}$ which depends on $\alpha$ has been
chosen as that $q$-value closest to the main maximum of $S_{00}(q)$.
Although the peak positions and also the intensity varies with $\alpha$ the
qualitative features do not depend on $\alpha$, at least for reasonable
chosen values for $\alpha$.

The $T^*$-dependence of the position $\omega_{op}$ and of the height $h_{op}$
of the op for $q\simeq 4.7$ is shown in Figure \ref{fig17n}a for $\phi_D=0.525$
along path D. The corresponding $\phi$ dependence for the same $q$ value and 
$T^*_A=0.04$ along path A is given in Figure \ref{fig17n}b. $h_{op}$ follows
a linear $T^*$-dependence between $T^*=0.04$ (the lowest temperature we have
studied) and $T^*\approx 0.1$. In a
temperature region where $\omega_{op}$ has a minimum, $h_{op}$ shows a
crossover to a constant. As can be observed from Figure \ref{fig17n}b the
position $\omega_{op}$ and the height $h_{op}$ are almost constant below $
\phi_c(T^*_A=0.04)\simeq 0.3782$ and increase with $\phi$ above that value.

\section{Schematic Model}

\label{cap4}

In this section we will demonstrate that besides a high frequency peak the
occurrence of an additional microscopic low frequency peak can be obtained
from a schematic model. For dipolar hard spheres it has been shown in
section \ref{cap2} that this additional peak in $\chi_{00}^{\prime\prime}(q,
\omega)$ and $\phi_{00}^{\prime\prime}(q,\omega)$ 
has its origin in the orientational degrees of freedom and their
coupling to the center of mass motion. Therefore the schematic model must
contain at least two correlators $\phi_0(t)$ and $\phi_1(t)$, describing
center of mass and small $q$ orientational dynamics, respectively. The
topology of the phase diagram (Fig. \ref{fig3}) is related to the special
form of the memory kernels (Eq. (\ref{eq37}) and (\ref{eq37a})), i.e. $
m_{00}^{\alpha\alpha^{\prime}}$ does not only involve $
S_{00}(q_1,t)S_{00}(q_2,t)$ but contains also a feedback from the dipolar
motion, due to $S_{1m_1}(q_1,t)S_{1m_2}(q_2,t)$. In contrast $
m_{1m}^{\alpha\alpha^{\prime}}(q,t)$ only possesses a coupling between the
center of mass and the dipolar correlators, i.e. the latter can only become
non-ergodic if the center of mass correlator becomes frozen. A schematic
model which exactly describes this behavior is the Bosse-Krieger model \cite
{krieger86} 
\begin{eqnarray}
\label{eq43a}
\ddot{\phi}_a(t) & + & \Omega_a^2 \phi_a(t) + \nu_a \dot{\phi}_a(t) +
\\
\nonumber
 & + & \Omega_a^2 \int_0^t dt' m_a(t-t')\dot{\phi}_a(t') =  0,
\end{eqnarray}
$a  =  0,1$ with the memory kernels 
\begin{subeqnarray}
m_0(t)  &\equiv &  F_0(\phi_0(t),\phi_1(t))= \xi_1 \phi_0^2(t) + \xi_2 \phi_1^2(t)
\\
m_1(t)  &\equiv & F_1(\phi_0(t),\phi_1(t)) =\xi_3\phi_0(t)\phi_1(t)
\label{eq43}
\end{subeqnarray}
and initial conditions $\phi_a(0)=1$, $\dot{\phi}_a(t)=0$. To mimic the $
\phi $- and $T$-dependence of the dipolar hard spheres we choose 
\begin{equation}  \label{eq44}
\xi_1\equiv \phi,\quad\quad \xi_2\equiv x/T, \quad\quad \xi_3\equiv 1/T .
\end{equation}
This choice is motivated by the observation that the mode coupling between
center of mass and dipolar dynamics vanishes for $T\to \infty$, leaving only
the cage effect for simple hard spheres with ''packing fraction'' 
$\phi$. $x$ is
considered as an additional parameter. From (\ref{eq43a}) 
and (\ref{eq43}) we immediately get
the equations 
\begin{equation}  \label{eq45a}
\frac{f_a}{1-f_a}=F_a(f_0,f_1),\quad\quad a=1,2
\end{equation}
for the corresponding non-ergodicity parameters 
\begin{equation}  \label{eq46}
f_a=\lim_{t\to\infty} \phi_a(t).
\end{equation}
(\ref{eq45a}) has been solved for $x=0.15$. From this solution we get the
phase diagram for glass transitions presented in Figure \ref{fig17}.
Comparison with Figure \ref{fig3} reveals qualitative agreement between both
phase diagrams. We emphasize that this result holds only in a certain range
for $x$. The full $\xi_1-\xi_2-\xi_3$ or equivalently $\phi-T-x$ diagram was
explored in detail in Ref. \cite{goetze88}. Since $f_1=0$ in the 
glass phase II (see Figure \ref{fig17}), Eqs. (\ref{eq45a}) for $a=0$ reduce
to the $F_2$-model \cite{goetze91} for which the critical value for $\phi$
is $\phi_c^{F_2}=4$ and the
critical non-ergodicity parameter is $f_c^{F_2}=\frac{1}{2}$ (see horizontal
line in Figure \ref{fig18}).

Having found the same qualitative phase diagram as for dipolar hard spheres
we can now investigate the dynamics for the Bosse-Krieger model.
This has already been done in order to calculate the long time dynamics for
the $\alpha$-relaxation \cite{fuchs91b} and the $\alpha$- and $\beta$-
relaxation \cite{kaneko96}. An investigation which also includes the
microscopic frequency regime can be found in Ref. \cite{bosse89}. Figure 4b of
this Reference gives a hint for the existence of an extra peak between the
$\alpha$- and the hf-peak. But this phenomenon has not been discussed there.  

In Figure \ref{fig18} we present $\phi_0(t)$ for fixed frequency $\Omega_0=1$
and different values for $\Omega_1$. For $\Omega_1=\Omega_0$ we find the
typical two step relaxation process from $\phi_0(0)=1$ to a plateau at $
f_c\equiv f_c^{F_2}$ and a final decay to zero. With decreasing $\Omega_1$
we observe that a shoulder occurs which finally develops into a second
plateau above $f_c^{F_2}$ This can easily be understood. In case that $
\Omega_1$ becomes more and more \emph{soft}, with respect to $\Omega_0$, the
second term in $m_0(t)$ (cf. Eq. (\ref{eq43}a) varies slower and slower,
already on the microscopic time scale $t_0=\Omega_0^{-1}$. Therefore the
relaxation kernel $m_0(t)$ has a rather slowly varying
part. It is this part which generates the second plateau. Since the decay
from a pronounced plateau produces a relaxation peak in the corresponding
susceptibility spectrum we expect, besides the high frequency peak related
to $\Omega_0$ and the $\alpha$-peak, one more peak. This peak stems from the
decay from the upper plateau to that at $f_c^{F_2}$. Figures \ref{fig20n}
and \ref{fig21n} present $\chi_a^{\prime\prime}(\omega)$ for $\Omega_0=1$, $
\Omega_1=0.1$ and $\nu_0=\nu_1=0$ along path A' and B' (see Figure \ref
{fig17}), respectively. Starting in the ergodic phase not too close to the
transition line there is in $\chi_0^{\prime\prime}(\omega)$ 
the $\alpha$-peak (visible on the linear scale for $\epsilon<0$ and $n=1$,
only)  
 and the high frequency peak at $\omega\approx\Omega_0$, only. Approaching the
critical line, we 
observe that the high frequency peak shifts to higher frequencies, due to
''stiffening'' (see Ref. \cite{goetze00}), and the $\alpha$-peak to lower
frequencies. With this shift of the $\alpha$-peak, a third peak becomes
more pronounced at a position which is about the same as the position of the
main peak in $\chi''_1(\omega)$ and which scales with the bare frequency 
$\Omega_1$. 
This peak remains also in the glass phase.
To investigate the nature of that additional peak again we have plotted
$\phi''_a(\omega)$ in Figure \ref{fig22n} and \ref{fig23n} for path A' and
B', respectively. The hf-peak occurs already in the ''liquid'' phase
(at least for the lower ''temperature'') indicating a damped oscillatory dynamics.
The spectral shape of the hf-peak strongly resembles the shape of the
anomalous oscillation peak (AOP) found in Ref. \cite{goetze00} for a hard
sphere system, a one-and a two component schematic model. Therefore our
notation hf-peak should not be confused with the high-frequency sound peak in
Ref. \cite{goetze00}.
In contrast, the additional peak at $\omega\approx\Omega_1$ does not exist in
the ''liquid'' but in the ''glass'' phase, only. Hence, the character of the extra
peak changes from a relaxational to a damped oscillational type. Therefore 
the
occurrence and the qualitative features of the additional peak are quite 
 similar to the behavior found
for the system of dipolar hard spheres. Since $\Omega_1$ mimics the
orientational frequency $\Omega_{1m}(q)$, we have further support of the
orientational origin of the orientational peak found in 
$\chi_{00}^{\prime\prime}(q,\omega)$ and 
$\phi_{00}^{\prime\prime}(q,\omega)$ for 
dipolar hard spheres. In addition, the schematic model also
demonstrates that an additional microscopic low frequency peak can only
exist when the \emph{microscopic} dynamics for $\phi_1(t)$ is ''soft''
enough with respect to that of $\phi_0(t)$.

For completeness we mention that for the lower ''temperature'' (path A')
there exists in $\phi_0''(\omega)$ a further peak at $\omega\approx0.4$. This
peak becomes more prominent on a double-logarithmic representation  (see  inset of
Fig.\ref{fig22n}) and its position also scales with $\Omega_1$, and not with
$\Omega_0$. 
Hence, both peaks at $\omega\approx0.1$ and $\omega\approx0.4$ have their
origin in the ''orientational'' motion of $\phi_1(t)$ and its coupling to
$\phi_0(t)$ via mode coupling terms.
Performing a ''stiff-glass''-approximation \cite{goetze00} for the BK-model
(which has not yet been done) should allow to elucidate the peak structures
and the spectral shapes more in detail. 
Since this is not the purpose of the present paper this will be discussed
elsewhere. 

It is also instructive to neglect memory effects for $\phi_1(t)$, i.e. we
put $m_1(t)$ equal to zero. Then, (\ref{eq43a}) for $a=1$ describes a damped
harmonic oscillator with a feedback on $\phi_0(t)$ due to the second term in 
$m_0(t)$. For this ''reduced'' Bosse-Krieger
model we present  $\chi_0^{\prime\prime}(\omega)$ in Figure \ref{fig24n} for $\Omega_0=1$, $\Omega_1=0.1$, $
\nu_0=0$ and $\nu_1=0.1$. Note that we introduced a nonzero damping $\nu_1$ in
order to comply with the nonzero width of the main peak in $\chi_1''(\omega)$
for the original BK-model. As expected one observes an additional peak which
originates from the damped harmonic motion of $\phi_1(t)$. Such an
observation has already been made earlier for another type of schematic
model \cite{franosch98,das99}. Comparing  $\chi_0''(\omega)$ from  Figure 
\ref{fig21n} and $\phi_0''(\omega)$ from Figure \ref{fig23n} with 
 $\chi_0''(\omega)$ and $\phi_0''(\omega)$ in Figure
 \ref{fig24n} shows that the memory effects for $\phi_1(t)$ do not influence 
$\chi_{0}^{\prime\prime}(\omega)$ and $\phi_{0}^{\prime\prime}(\omega)$ 
qualitatively.

\section{Discussion and Conclusions}

\label{cap5}

Recently we have extended the mode coupling theory (MCT) for simple or
binary  \cite{goetze91,goetze92,schilling94,cummins99} to molecular
liquids \cite{schilling97,fabbian99,letz00}. This has been done within a
molecular representation, which separates translational from orientational
degrees of freedom. This molecular theory has already been applied and
tested for diatomic molecules \cite{theis98b,schilling00,winkler00} and
water molecules with SPC/E potential \cite{fabbian99,theis00}. These tests
were restricted to the comparison of the non-ergodicity parameters and
critical amplitudes (for diatomic molecules, only) from molecular MCT with
the corresponding quantities from a MD-simulation. 
A satisfactory agreement between MCT and MD-simulation has been found
So far no dynamical
results were determined from mode coupling theory for \emph{molecular
liquids}.
The only solution of the time-dependent MCT-equations for a system with
orientational degrees of freedom has been obtained recently for
the orientational
correlators of a single dumbell in an isotropic liquid \cite
{chong00,goetze00b}. In the present paper we have solved the time-dependent
molecular MCT equations for a system of dipolar hard spheres. This is one of
the simplest systems involving translational and orientational degrees of
freedom. In addition it has the advantage, that the static correlators which
are the input quantities for the MCT-equations are known analytically within
some approximations. These approximations make the tensorial correlators in
the $q$-frame $\phi_{lm,l^{\prime}m^{\prime}}(\vec{q}=\vec{q}_0,t)=
\phi_{lm}(q,t)\delta_{ll^{\prime}} \delta_{mm^{\prime}}$ diagonal in $l$ and 
$l^{\prime}$. In addition they lead to a restriction of  $\phi_{lm}(q,t)$ to $
l=0,1 $ only. This means that our calculations yield information on center
of mass and dipolar dynamics.

The major result we have found in the microscopic frequency regime is the
existence of an orientational peak at $\omega_{op}($liquid$)\approx
\hat{\omega}_{op}($glass$)$ about one decade below the high
frequency peak. This orientational peak exists above and below the glass transition
lines in the two dimensional phase diagram of a dimensionless temperature $
T^*$ and packing fraction $\phi$ provided that the $\alpha$-peak position $
\omega_\alpha$ is much smaller than $\omega_{op}$.
On the liquid side it exists in $\chi_{00}''(q,\omega)$, the compressibility
spectrum but not in  $\phi_{00}''(q,\omega)$ which proves its pure
relaxational character.
However, crossing the glass transition line this peak also appears in the
correlation spectrum  $\phi_{00}''(q,\omega)$ if $T^*$ and $\phi$ becomes,
respectively, small and large enough.
This means that the orientational peak exists on {\em both} sides of the glass
transition lines and changes from a relaxational to a damped oscillational
type of excitation under transformation of the liquid to the glass.
The manner how the op develops in  $\chi_{00}''(q,\omega)$ as a function of
$\phi$ (see the double log representation in Fig. \ref{fig6}) resembles 
 at least qualitatively, the
behavior of the extra peak in the susceptibility spectra for e.g. 
Salol and Orthoterphenyl \cite{cummins93,cummins97,cummins98} 
at $\omega\approx 300$ GHz under variation of the
temperature.

Since the position $\omega_{op}$ is almost independent
on $q$ it is related to a localized, non-propagating mode. In addition, the
isotope-effect, i.e. $\omega_{op} \propto I^{-1/2}$, clearly proves its
orientational origin. Since the bare frequencies $\Omega_{lm}(q)$ for $l=1$
scale with $I^{-1/2}$ for $q=0$, only (cf. Eq. (\ref{eq42})), the orientational peak
must be generated by orientational modes with $q\approx0$. Why do so
long-waved orientational modes play a crucial role? An answer follows from
the static correlator $S_{11}(q)$. In contrast to $S_{00}(q)$ and $S_{10}(q)$
it exhibits a dominant peak with width $\Delta q$ at $q=0$ and decays
rapidly to one for $q>\Delta q$ (see Fig. \ref{fig1}). With decreasing
temperature its peak height increases and $\Delta q$ decreases. The increase
of $S_{11}(q\approx 0)$ has two implications: 

(i) since the glass transition
in MCT is driven by the growing of the main peaks of the static
correlators \footnote{That the underlying mechanism leading to an ideal glass
transition can be much more sophisticated, i.e. not only related to the
increase of e.g. one peak in $S_{00}(q)$ has been recently demonstrated for a
system of hard spheres with attractive interactions
\cite{fabbian98,zaccarelli01,dawson01,bergenholtz99}},
it is the increase of $S_{11}(q\approx 0)$ and the coupling of the
orientational mode $\rho_{11}(\vec{q},t)$ to $\rho_{00}(\vec{q},t)$, the
translational one, which enhances the tendency for glass formation.
This type of glass transition mechanism was already described within the
framework of MCT for {\em crystalline} systems.
For orientational glasses \cite{michel88,bostren91} and strictly periodic lattices
\cite{aksenov87,schilling97b} 
which undergo a second order equilibrium phase transition the
increase of the critical fluctuations already lead to an ideal glass
transition above the equilibrium transition temperature.
The increase of the critical fluctuations is accompanied by the softening of
the critical mode.

\begin{figure}
	\centerline{\rotatebox{-0}{\resizebox{9cm}{!}
	{\includegraphics{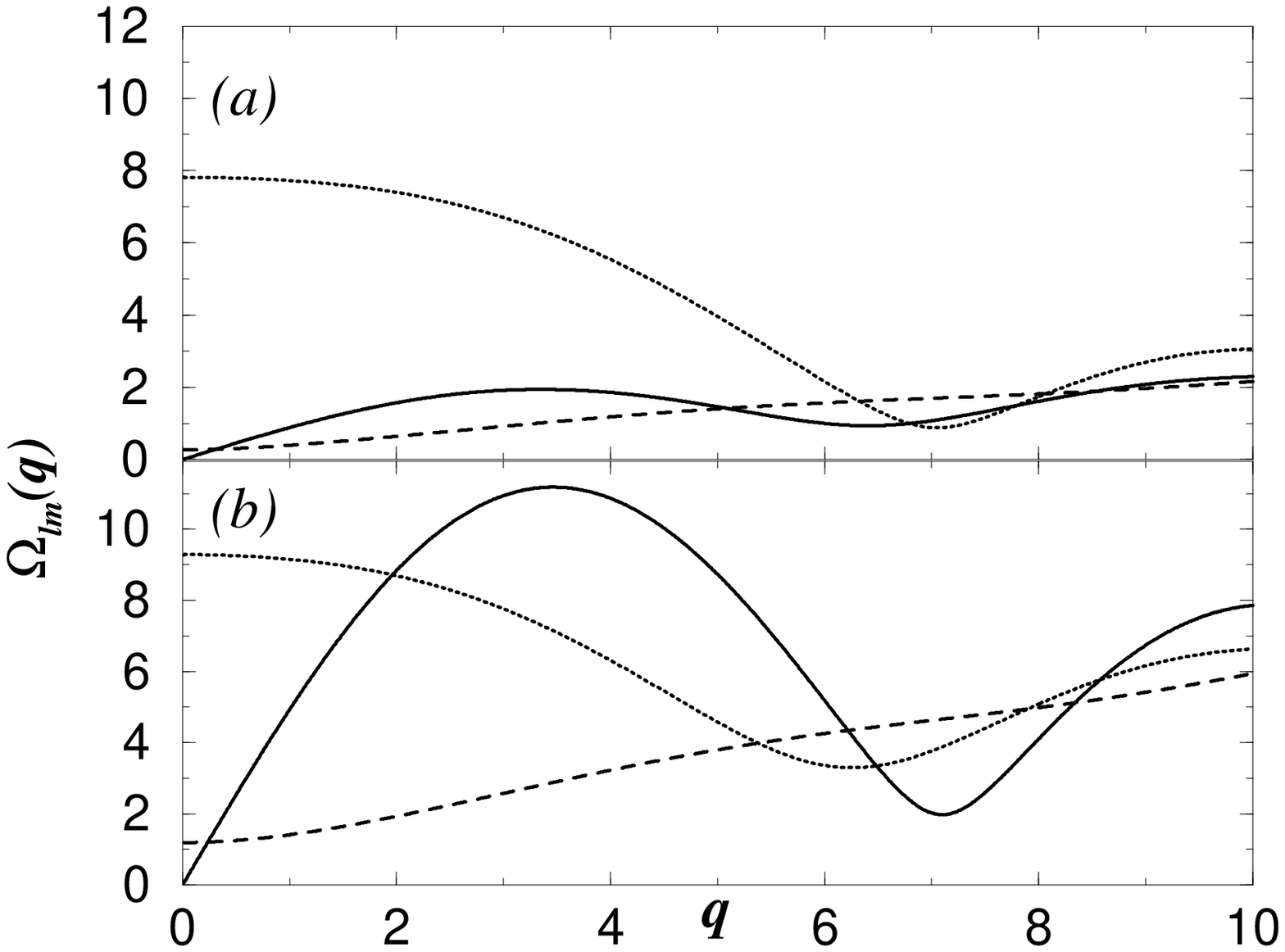}}}}
	\caption{$q$-dependence of the microscopic frequencies 
	$\Omega_{lm}(q)$ for $(l,m)=
	(0,0)$ (solid), $(1,0)$ (dotted) and $(1,1)$ (dashed)
	(a) $\phi=0.381$, $T^*=0.04$ and (b) $\phi=0.53$, $T^*=0.3$, indicated
	in Figure \ref{fig3} by the full circles}
\label{fig4}
	\centerline{\rotatebox{-0}{\resizebox{9cm}{!}
	{\includegraphics{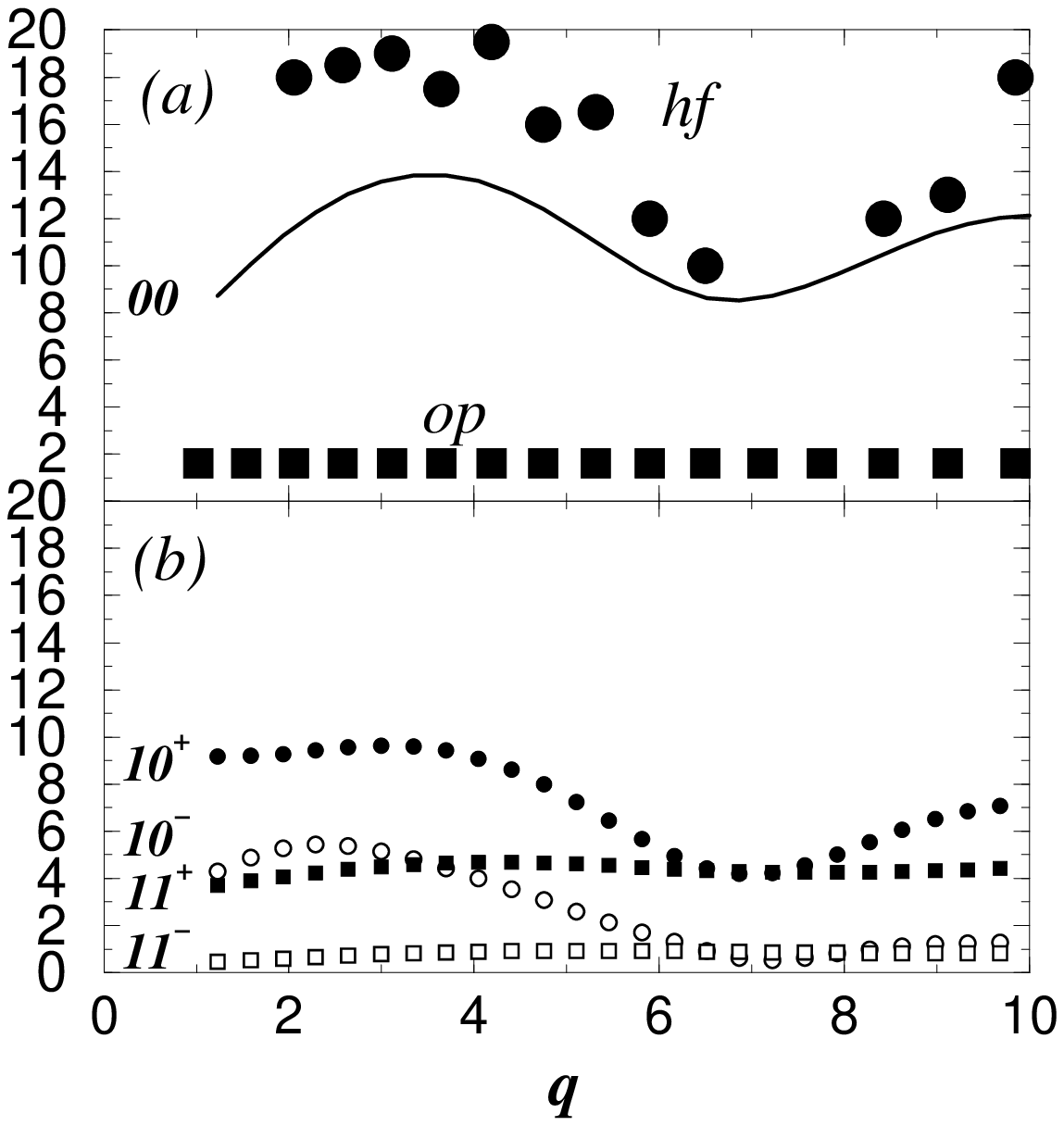}}}}
	\caption{$q$-dependence of the renormalized microscopic frequencies 
	for $\phi=0.53$, $T^*=0.04$: 
	(a) $\hat{\Omega}_{00}(q)$ is shown as solid line; the position
	$\hat{\omega}_{hf}$ and $\hat{\omega}_{op}$ obtained from
	$\phi''_{00}(q,\omega)$ are presented by the full circles and spheres,
	respectively
	(b) $\hat{\Omega}^\pm_{10}(q)$ and $\hat{\Omega}^\pm_{11}(q)$ are
	shown as circles and spheres, respectively.  }
\label{fig4a}
\end{figure}

\begin{figure}
	\centerline{\rotatebox{-0}{\resizebox{9cm}{!}
	{\includegraphics{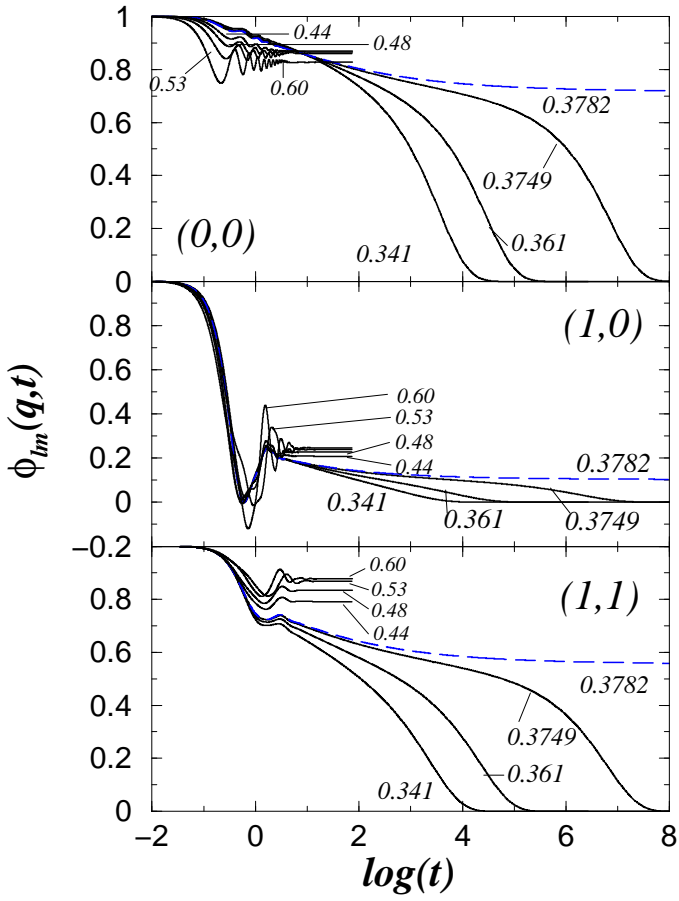}}}}
	\caption{lin-log plot of $\phi_{lm}(q,t)$ at $q\simeq 4.7$, 
	$T^*_A=0.04$ and $\phi$ along path A
	in Figure \ref{fig3}. 
	The dashed line corresponds to the results at
	$\phi \simeq \phi_c\simeq 0.3782$}
\label{fig5}
\end{figure}

\begin{figure}
	\centerline{\rotatebox{-0}{\resizebox{9cm}{!}
	{\includegraphics{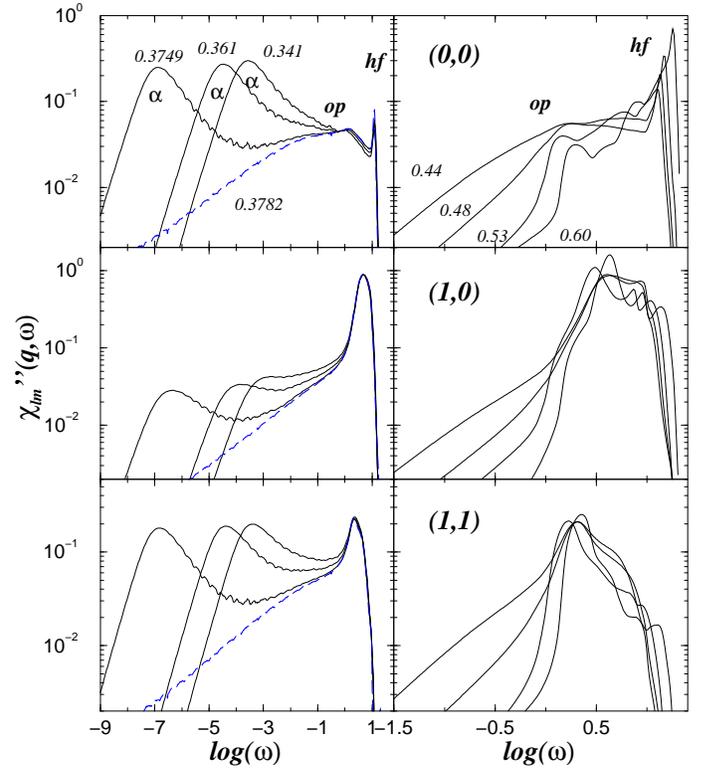}}}}
	\caption{log-log plot of $\chi''_{lm}(q,\omega)$ at $q\simeq 4.7$, 
	$T^*_A=0.04$ and $\phi$ along path A
	in Figure \ref{fig3}: liquid (left column) and glass (right column). 
	The dashed line corresponds to the results at
	$\phi \simeq \phi_c\simeq 0.3782$. $\phi$-values for
	$\chi''_{10}(q,\omega)$ and $\chi''_{11}(q,\omega)$ are the same as
	for $\chi''_{00}(q,\omega)$}
\label{fig6}
\end{figure}

(ii) since $\Omega_{11}(q) \propto 1/\sqrt{S_{11}(q)}$ (cf. Eq. (\ref{eq42}
)) the increase of $S_{11}(q\approx 0)$ implies a softening of the
orientational mode for $q\approx 0$ (see also Fig. \ref{fig4}). It is this 
\emph{softening} of the orientational mode with $q\approx 0$ and $l=1$, $m=1$
which results in a frequency scale separation between this orientational
mode and the translational one ($l=0$, $m=0$) with $q\approx q_{max}$.
Therefore two peaks, the orientational and the hf-peak, become visible.
The connection of $\hat{\omega}_{op}(q)$ with the $(1,1)$-mode is also
supported by the fact that
$\hat{\omega}_{op}(q)\approx\hat{\Omega}^\pm_{11}(q)$ (see Fig. \ref{fig4a}).   

The narrowing of the peak in $S_{11}(q)$ at $q=0$ also has an interesting
implication. Since $1/\Delta q$ is a measure of a correlation length, the
narrowing implies a growth of an orientational order. Although $1/\Delta q$
does \emph{not} diverge, it may be significantly larger than a few
Angstr\"om, i.e there may exist a \emph{medium range orientational} order.
For instance, for $\phi=0.381$ and $T^*=0.04$ the correlation length is
about six diameters (cf. Figure \ref{fig1}). The fact that the increase of $
S_{11}(q\approx 0)$ is accompanied by the narrowing of the peak in $
S_{11}(q) $ at $q=0$ proves the existence of a correlation between the
occurrence of the orientational peak and medium range orientational order. This is
an interesting observation insofar that such a correlation has already been
predicted \cite{elliot92,sokolov92,sokolov93,foley95,fayos96}, although this
does not seem to be a universal feature \cite{boerjesson93}. 
We emphasize that we do not
consider this type of behavior as an exceptional case which exists for
dipolar hard spheres, only. For example for a system of hard ellipsoids
which may exhibit medium ranged nematic order (see Ref. \cite{letzschill99})
similar results are expected.

Besides the op, we have found in $\chi''_{00}(q,\omega)$ and
$\phi''_{00}(q,\omega)$ above and below the glass transition a high frequency
peak at $\omega_{hf}(q)$. It exists in the investigated range $0.51<q<40$ except
in an interval centered around $q_{max}$, the position of the main peak in $
S_{00}(q)$. 
$\omega_{hf}(q)$ is shown in Figure \ref{fig4a} for $q\leq 2$. 
Despite an inaccuracy of about 10\ (due to the $q$-discretization) its
$q$-dependence is in phase with the $q$-variation of the renormalized
frequency $\hat{\Omega}_{00}(q)$.
These qualitative features are similar to
the type of the high frequency peak found for a system of hard spheres
\cite{goetze00}. 
Since the accuracy of $\omega_{hf}(q)$ becomes even worse for $q\to 0$ it is
not possible to detect the expected linear $q$-dependence for $q\to0$.
Accordingly, we can not yet prove that the hf-peak for DHS corresponds to
the high frequency sound, as found for the system of hard spheres
\cite{goetze00}. 
In \cite{goetze00} 
 it has also been shown that an additional peak, which was
called ''anomalous oscillation peak'' (AOP), appears deep in the glass
phase. Its origin lies in the harmonic motion of the particles in their
cages \cite{goetze00}. Whether or not the hump between the op and the hf-peak
we have observed in $
\phi_{00}^{\prime\prime}(q,\omega)$ deep in the glass (see Fig. \ref{fig10n}
and \ref{fig15n}) corresponds to the AOP is not clear. Again due to the
restriction 
to $30$ values for $q$ and also a smaller number of discrete values for $t$
compared to Ref. \cite{goetze00} we can not give detailed \emph{
quantitative} information on, e.g. the $q$-dependence of the position and
width of that hump.
On the other
hand the high frequency wing of the orientational peak may interfere with the AOP,
thus complicating the analysis of the latter.

\begin{figure}
	\centerline{\rotatebox{-0}{\resizebox{9cm}{!}
	{\includegraphics{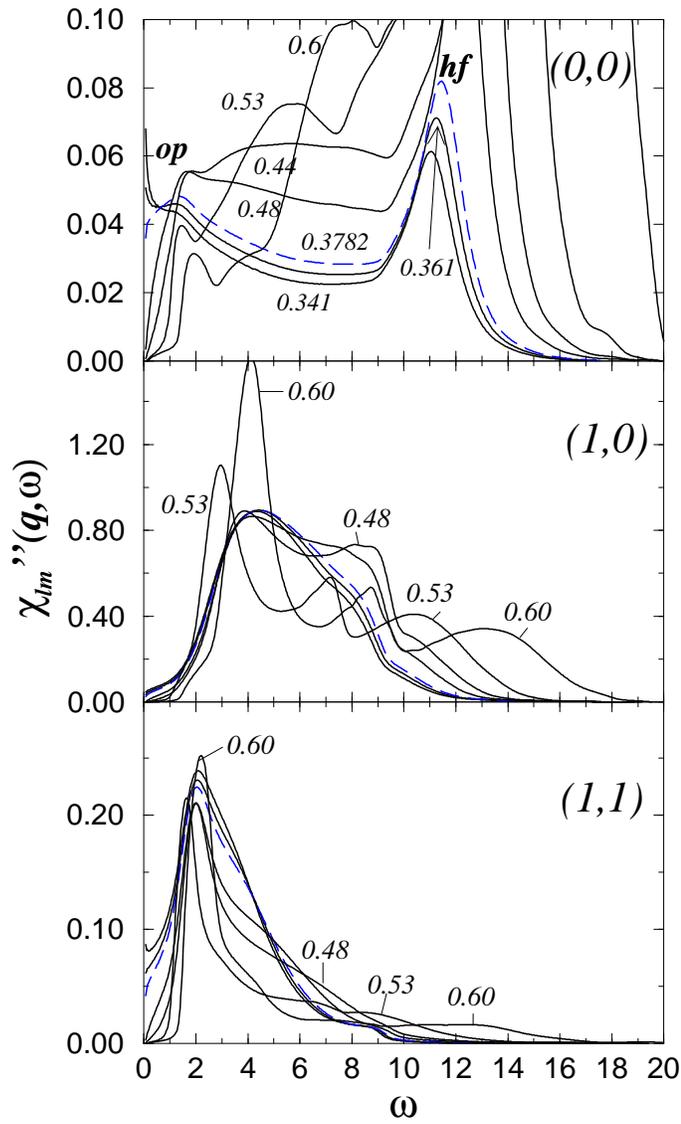}}}}
	\caption{$\chi''_{lm}(q,\omega)$ on linear scales at $q\simeq 4.7$, 
	$T^*_A=0.04$ and $\phi$ along path A
	in Figure \ref{fig3}. 
	The dashed line corresponds to the results at
	$\phi \simeq \phi_c\simeq 0.3782$}
\label{fig7n}
\end{figure}

\begin{figure}
	\centerline{\rotatebox{-0}{\resizebox{9cm}{!}
	{\includegraphics{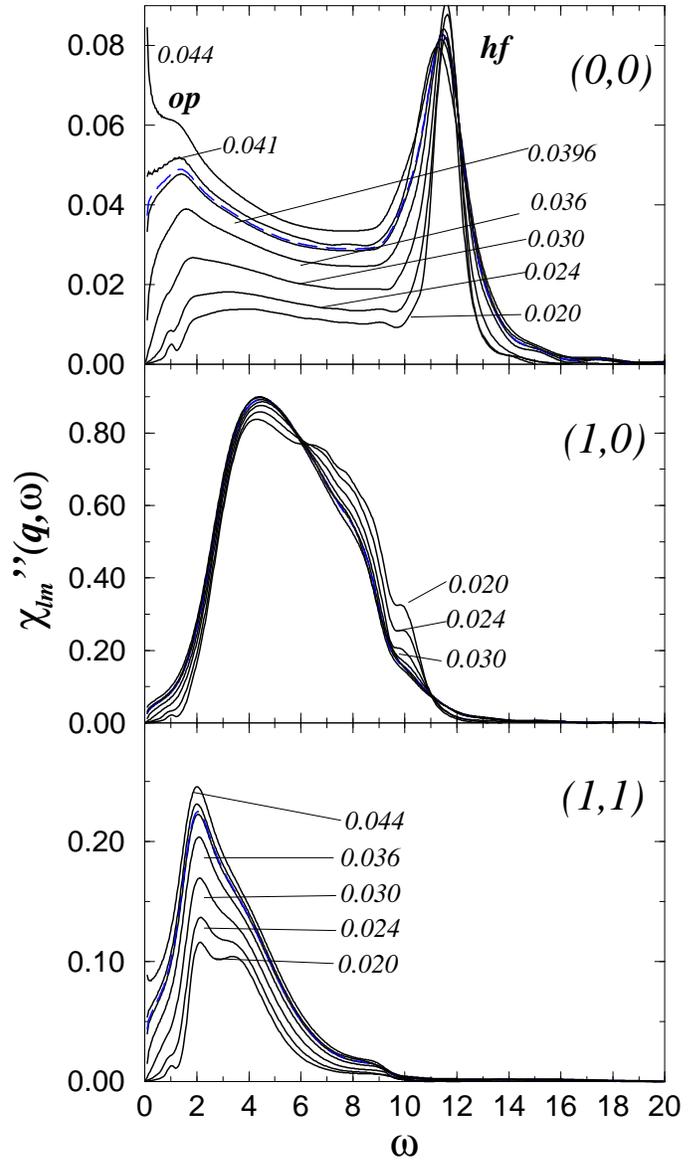}}}}
	\caption{ $\chi''_{lm}(q,\omega)$ on linear scales at $q\simeq 4.7$, 
	$\phi_C=0.3786$ and $T^*$  along path C.
	The dashed line corresponds to the result at
	$T^* \simeq T^*_c\simeq 0.04$}
\label{fig8n}
\end{figure}

\begin{figure}
	\centerline{\rotatebox{-0}{\resizebox{9cm}{!}
	{\includegraphics{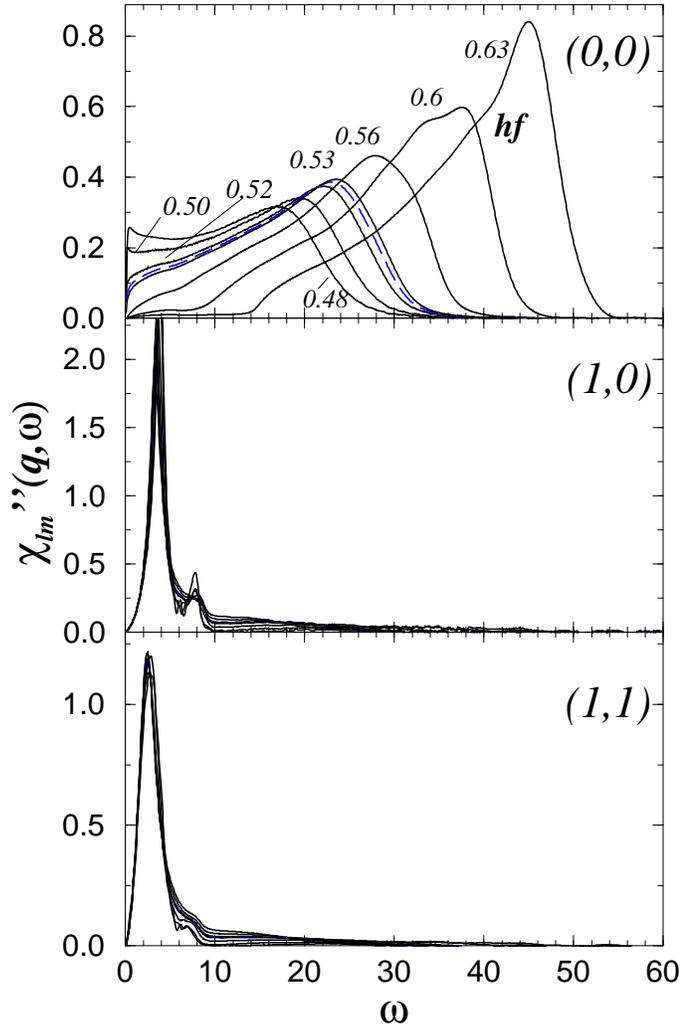}}}}
	\caption{ $\chi''_{lm}(q,\omega)$ on linear scales at $q\simeq 4.7$, 
	$T^*_B=0.3$ and $\phi$  along path B.
	The dashed line corresponds to the result at
	$\phi \simeq \phi_c\simeq 0.5265$. Because $\chi''_{1m}(q,\omega)$
	does not vary much with $\phi$ we have not labelled the various curves
	with $\phi$}
\label{fig9n}
\end{figure}

\begin{figure}
	\centerline{\rotatebox{-0}{\resizebox{9cm}{!}
	{\includegraphics{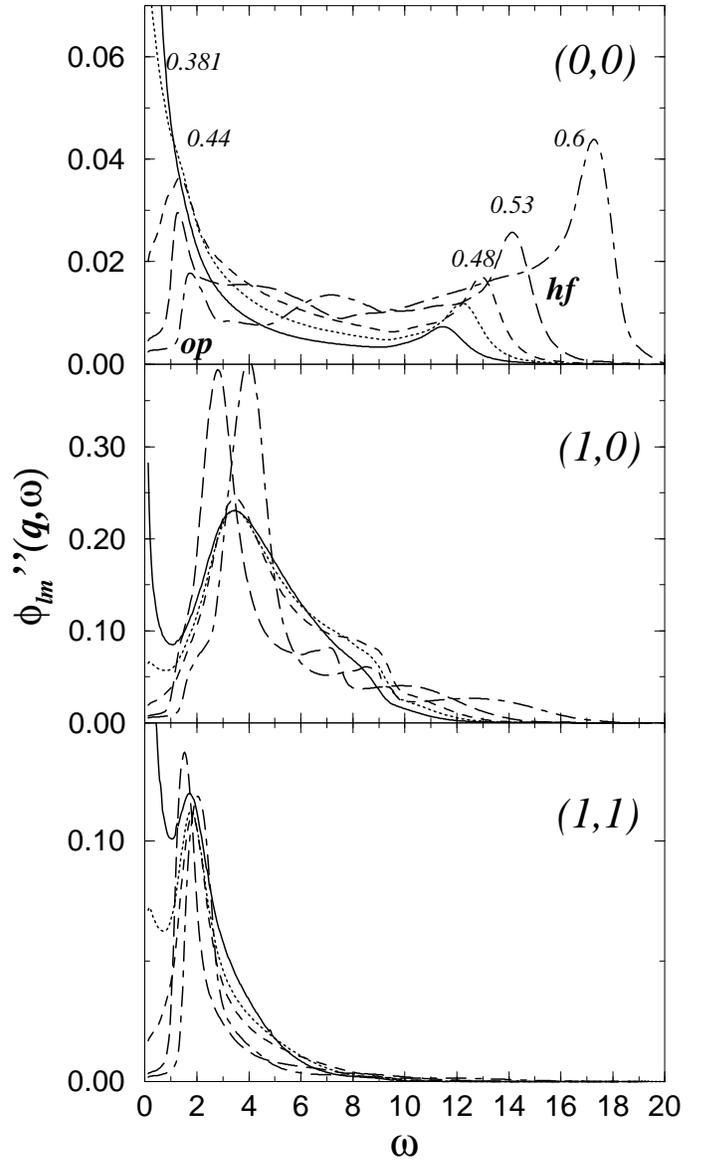}}}}
	\caption{$\phi''_{lm}(q\simeq 4.7,\omega)$ on linear scales
	in the glass phase along path
	A. I.e. $T^*=0.04$ and 
			$\phi=0.381$ (solid),
			$\phi=0.44$ (dotted),
			$\phi=0.48$ (dashed),
			$\phi=0.53$ (long-dashed) and
			$\phi=0.6$ (dot-dashed)}
\label{fig10n}
\end{figure}

\begin{figure}
	\centerline{\rotatebox{-0}{\resizebox{9cm}{!}
	{\includegraphics{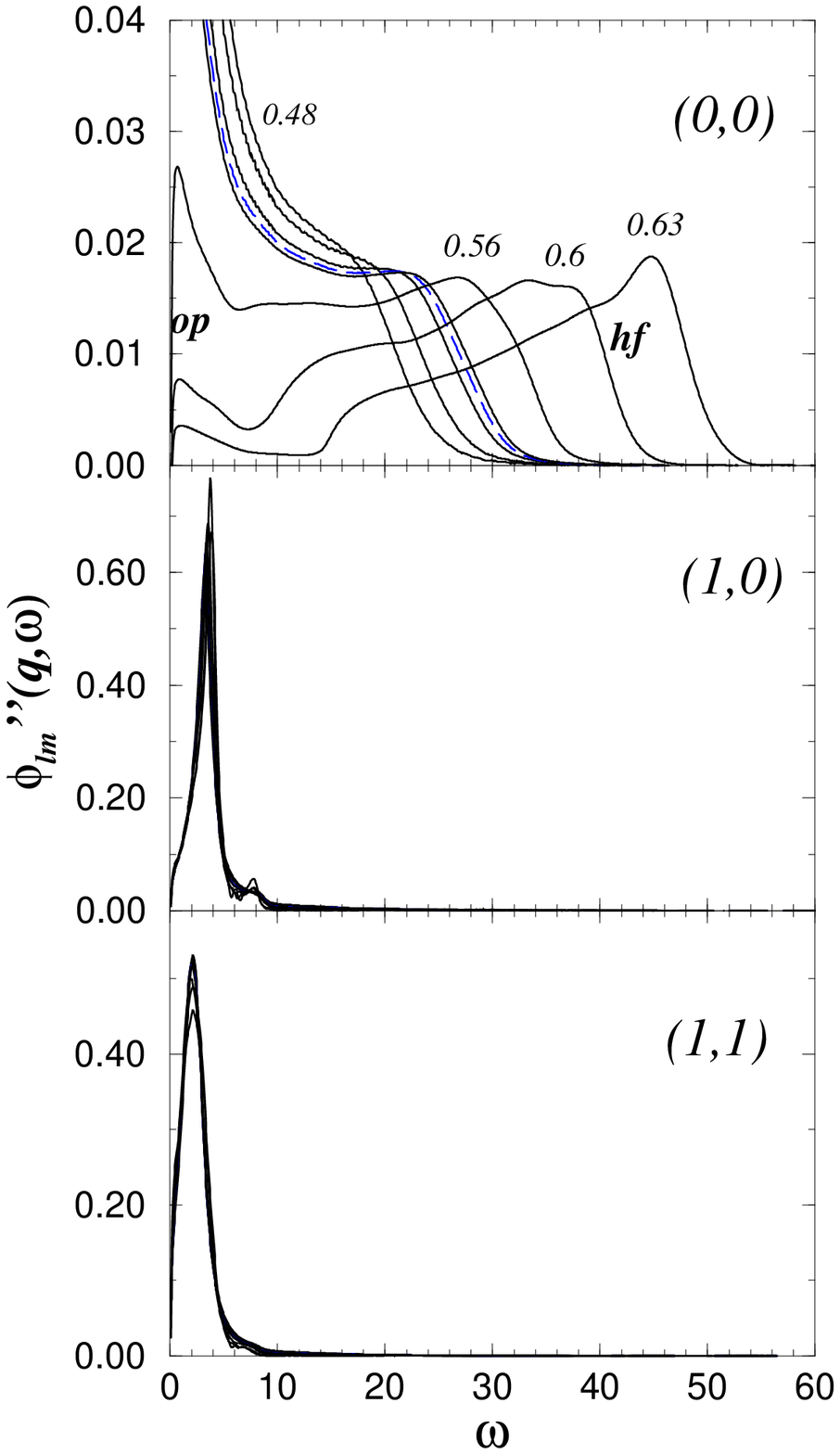}}}}
	\caption{$\phi''_{lm}(q\simeq4.7,\omega)$ on linear scales for
	$T^*=0.30$ and $\phi$ along path B. The dashed line corresponds to
	the result at $\phi \simeq \phi_c\simeq 0.5265$. Because
	$\phi''_{1m}(q,\omega)$ 
	does not vary much with $\phi$ we have not labelled the various curves
	with $\phi$}
\label{fig11n}
\end{figure}

\begin{figure}
	\centerline{\rotatebox{-0}{\resizebox{9cm}{!}
	{\includegraphics{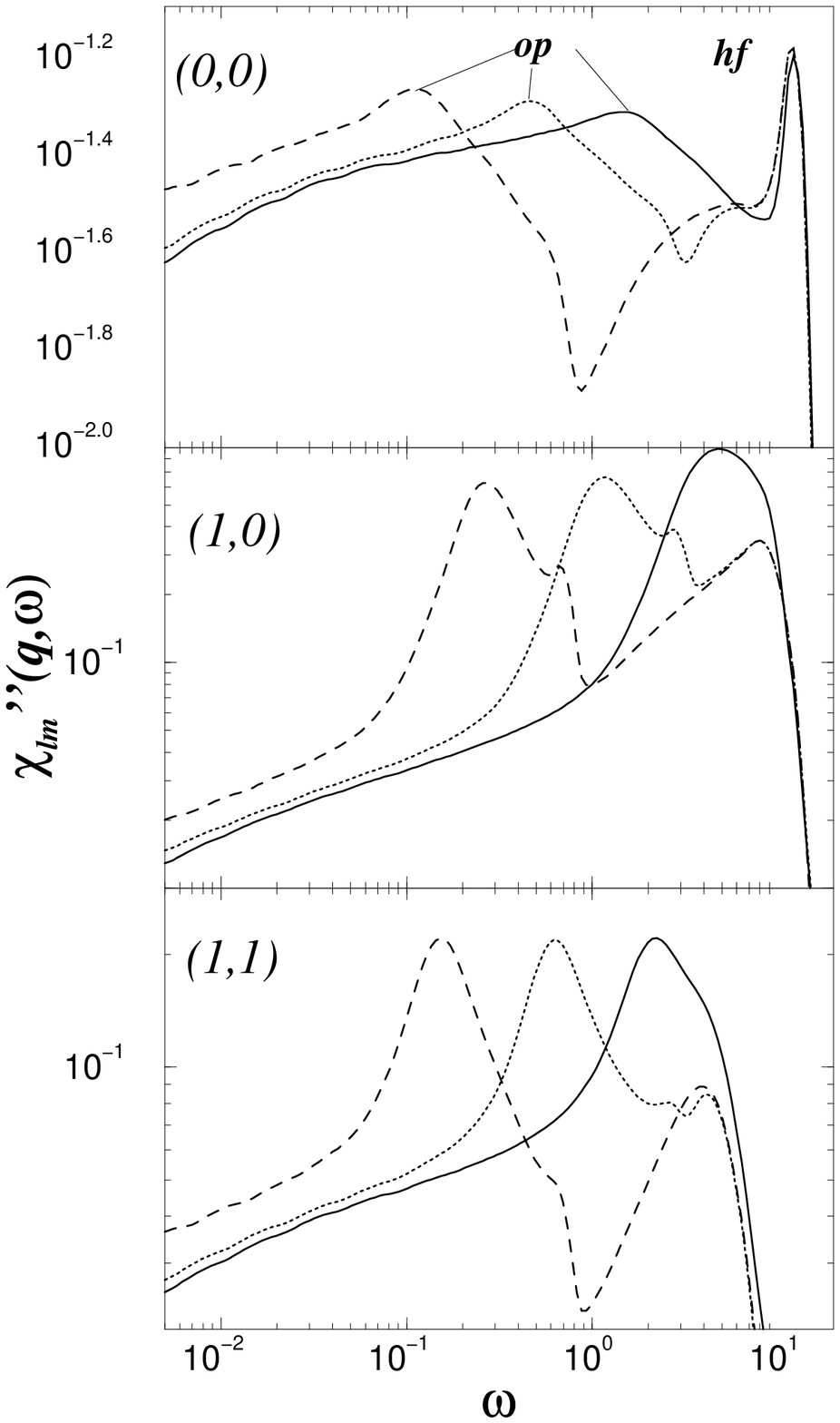}}}}
	\caption{$\chi''_{lm}(q\simeq4.7,\omega)$ at $\phi=0.381$, $T^*=0.04$
	for 
	$I=\frac{1}{10}$ (solid), $1$ (dotted) and $10$ (dashed)}
\label{fig12n}
\end{figure}

\begin{figure}
	\centerline{\rotatebox{-0}{\resizebox{9cm}{!}
	{\includegraphics{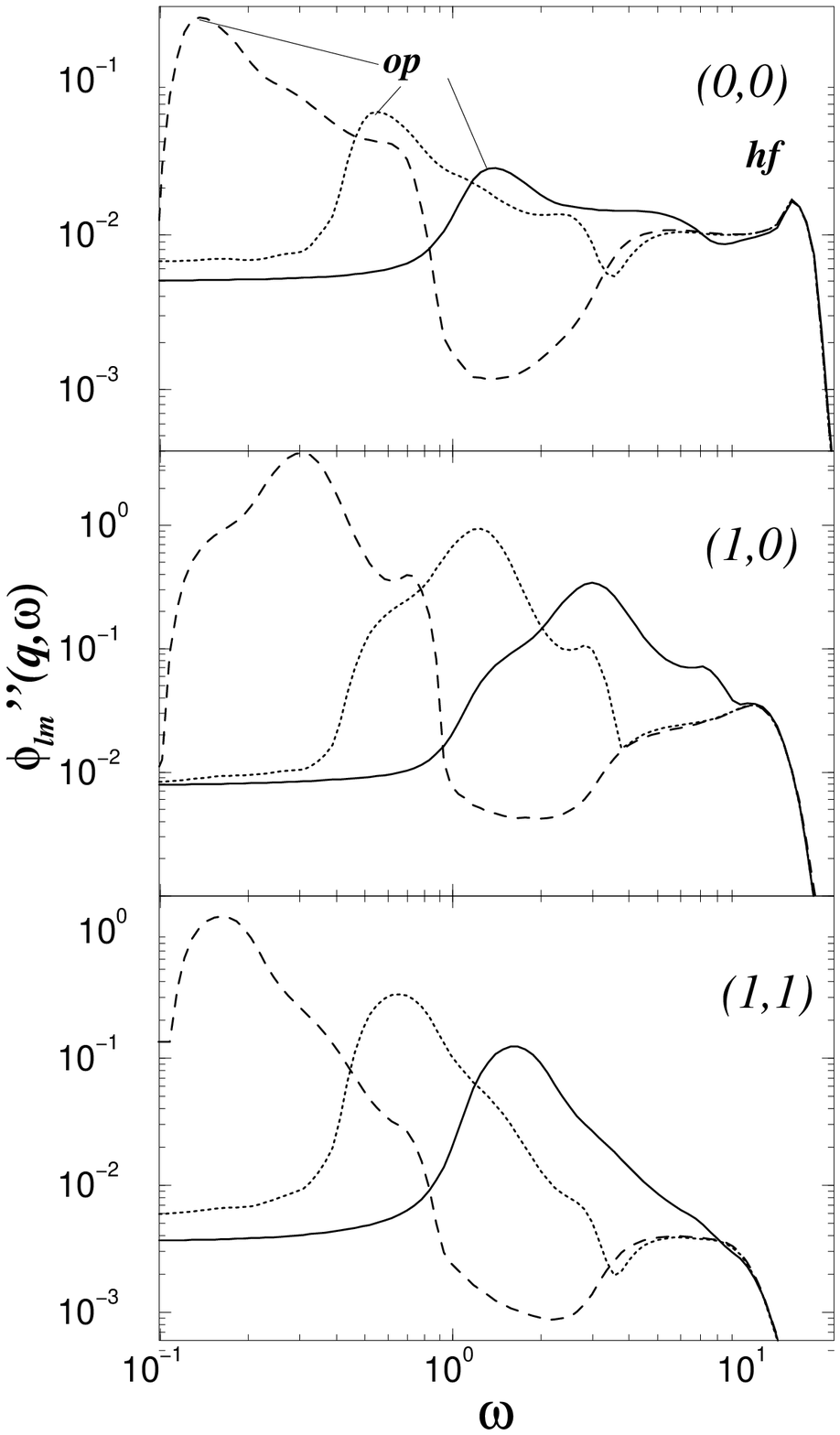}}}}
	\caption{$\phi''_{lm}(q\simeq4.7,\omega)$ at $\phi=0.53$, $T^*=0.04$ for
	$I=\frac{1}{10}$ (solid), $1$ (dotted) and $10$ (dashed)} 
\label{fig13n}
\end{figure}

\begin{figure}
	\centerline{\rotatebox{-0}{\resizebox{9cm}{!}
	{\includegraphics{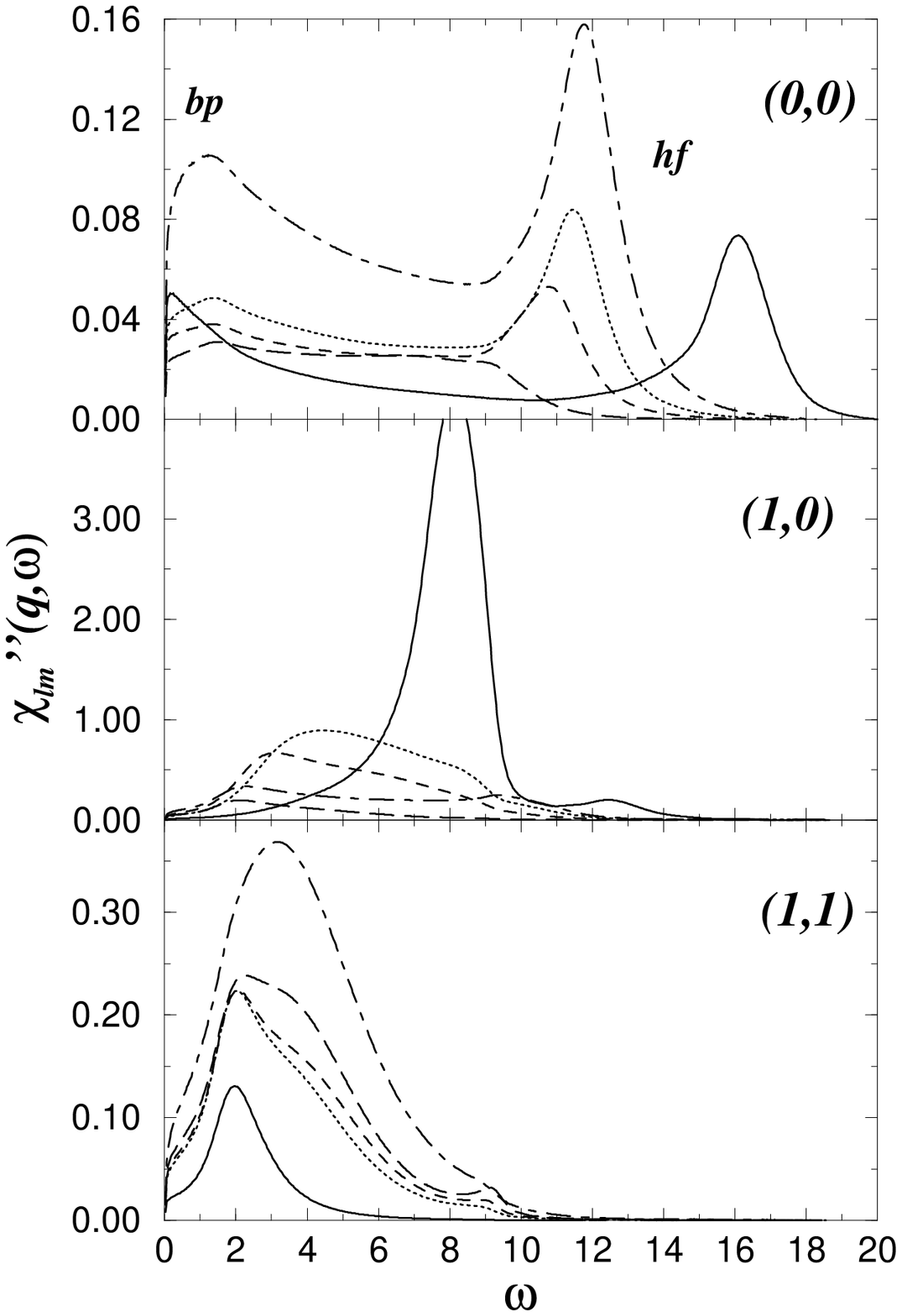}}}}
	\caption{$q$-dependence of $\chi''_{lm}(q,\omega)$ at $\phi=0.381$,
	$T^*=0.04$ for 
	$q\simeq 2.0$ (solid), $q\simeq 4.7$ (dotted),
	$q\simeq 5.3$ (dashed), $q\simeq 6.5$ (long dashed) 
	and  $q\simeq 9.8$ (dot dashed)}
\label{fig14n}
\end{figure}

\begin{figure}
	\centerline{\rotatebox{-0}{\resizebox{9cm}{!}
	{\includegraphics{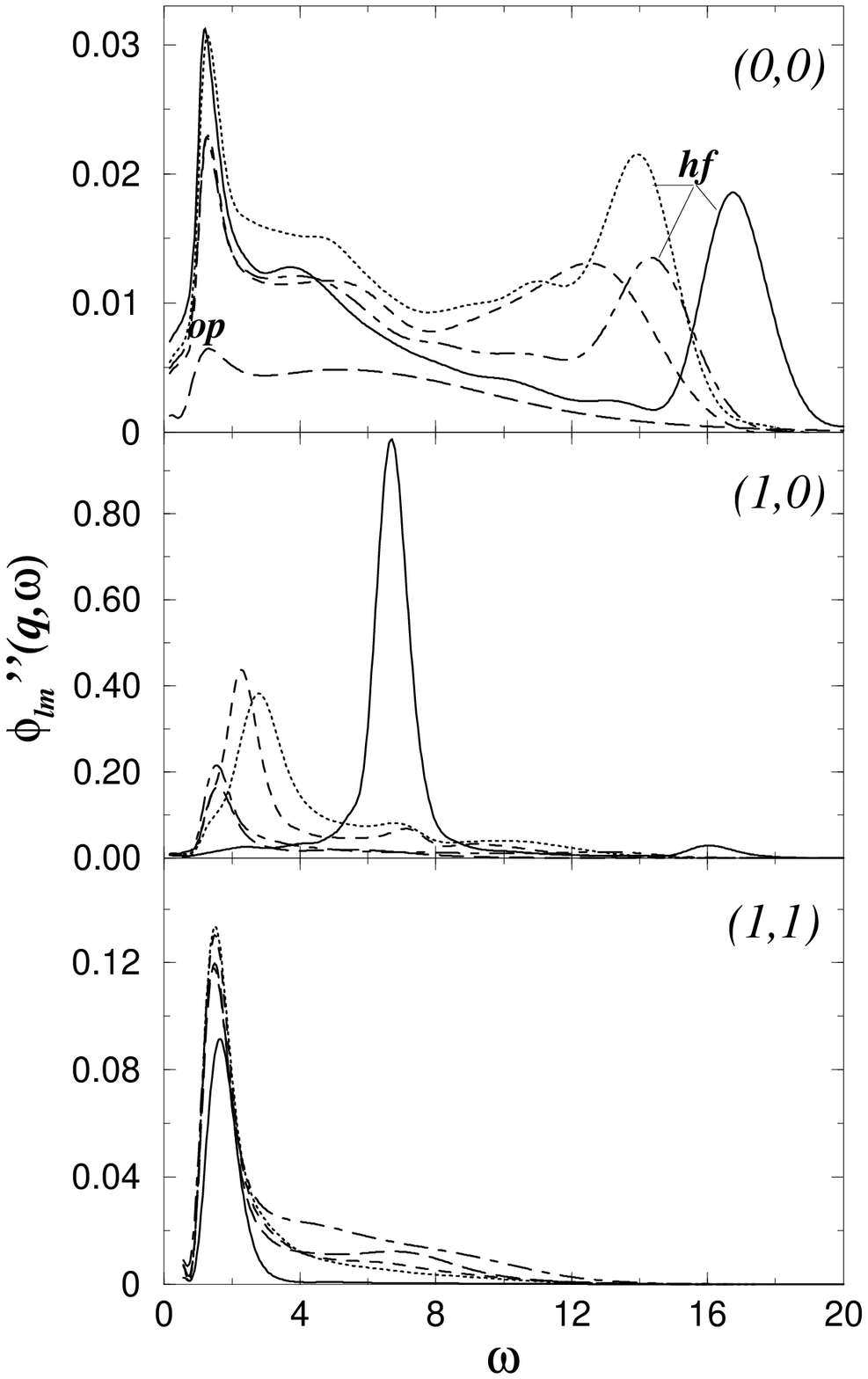}}}}
	\caption{$q$-dependence of $\phi''_{lm}(q,\omega)$ at 
	$\phi=0.53$, $T^*=0.04$ for
	$q\simeq 2.0$ (solid), $q\simeq 4.7$ (dotted),
	$q\simeq 5.3$ (dashed), $q\simeq 6.5$ (long dashed) 
	and  $q\simeq 9.8$ (dot dashed)}
\label{fig15n}
\end{figure}

\begin{figure}
	\centerline{\rotatebox{-0}{\resizebox{8cm}{!}
	{\includegraphics{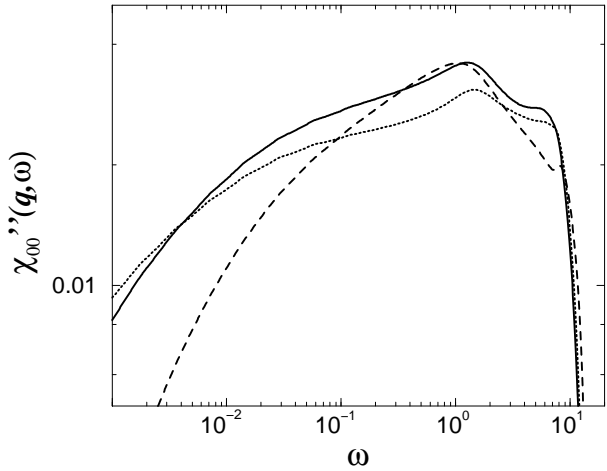}}}}
	\caption{$\alpha$ - dependence of
        $\chi''_{00}(\bar{q}_{max},\omega)$ at  
	$\phi=0.381$, $T^*=0.04$ for $\alpha=0.06$ (dashed), $0.065$
	(solid) and $0.068$ (dotted)}
\label{fig16n}
\end{figure}

\begin{figure}
	\centerline{\rotatebox{-0}{\resizebox{9cm}{!}
	{\includegraphics{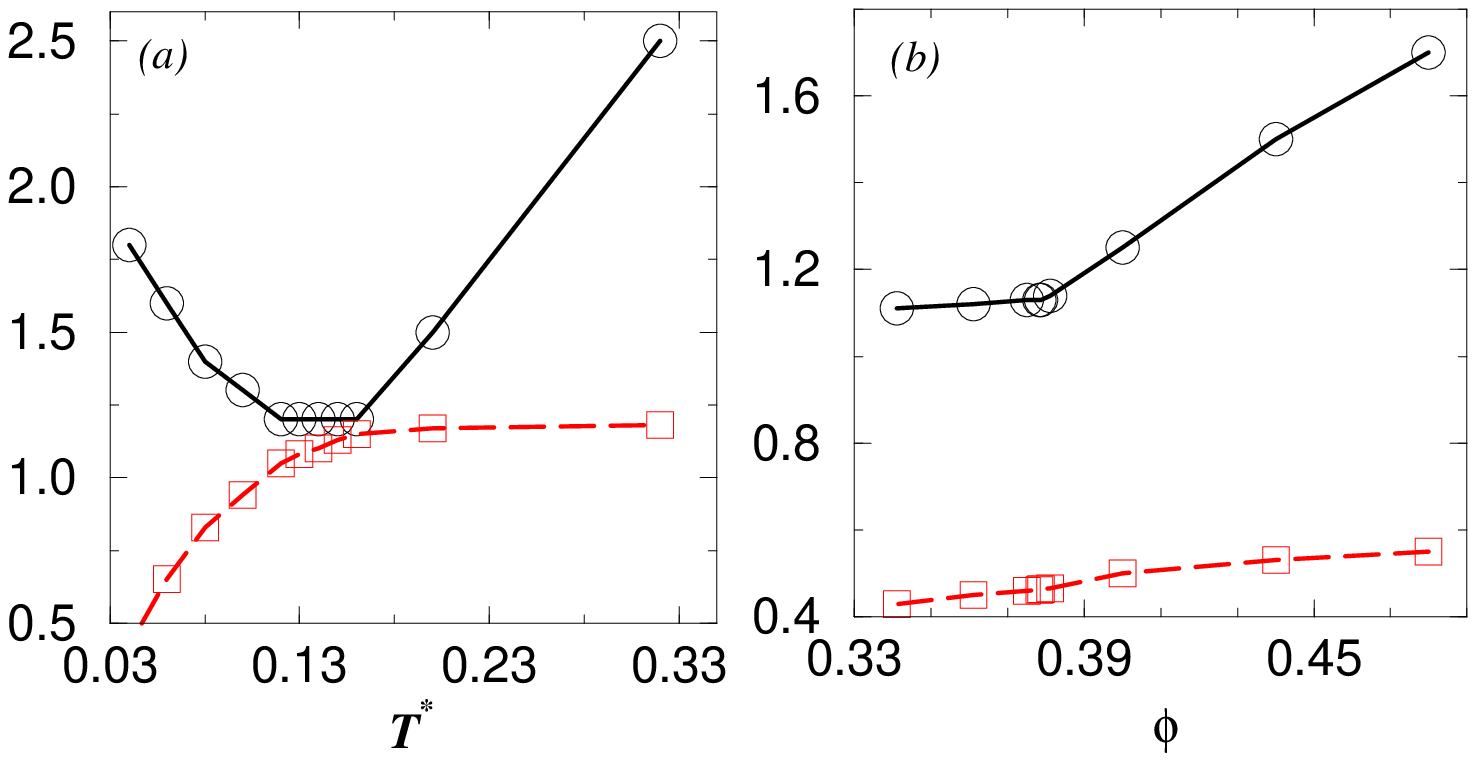}}}}
	\caption{
	(a) Temperature dependence of the position 
	$\omega_{op}$ ($\circ$) and height $h_{op}$ ($\Box$) (multiplied by
	$10$) of the orientational
	peak for $q\simeq 4.7$ and $\phi_D=0.525$ along path D.
	(b) Dependence on the packing fraction  of the position 
	$\omega_{op}$ ($\circ$) and height $h_{op}$ ($\Box$) (multiplied by
	$10$) of the orientational
	peak for $q\simeq 4.7$ and $T^*_A=0.04$  along path A}
\label{fig17n}
	\centerline{\rotatebox{-0}{\resizebox{7cm}{!}
	{\includegraphics{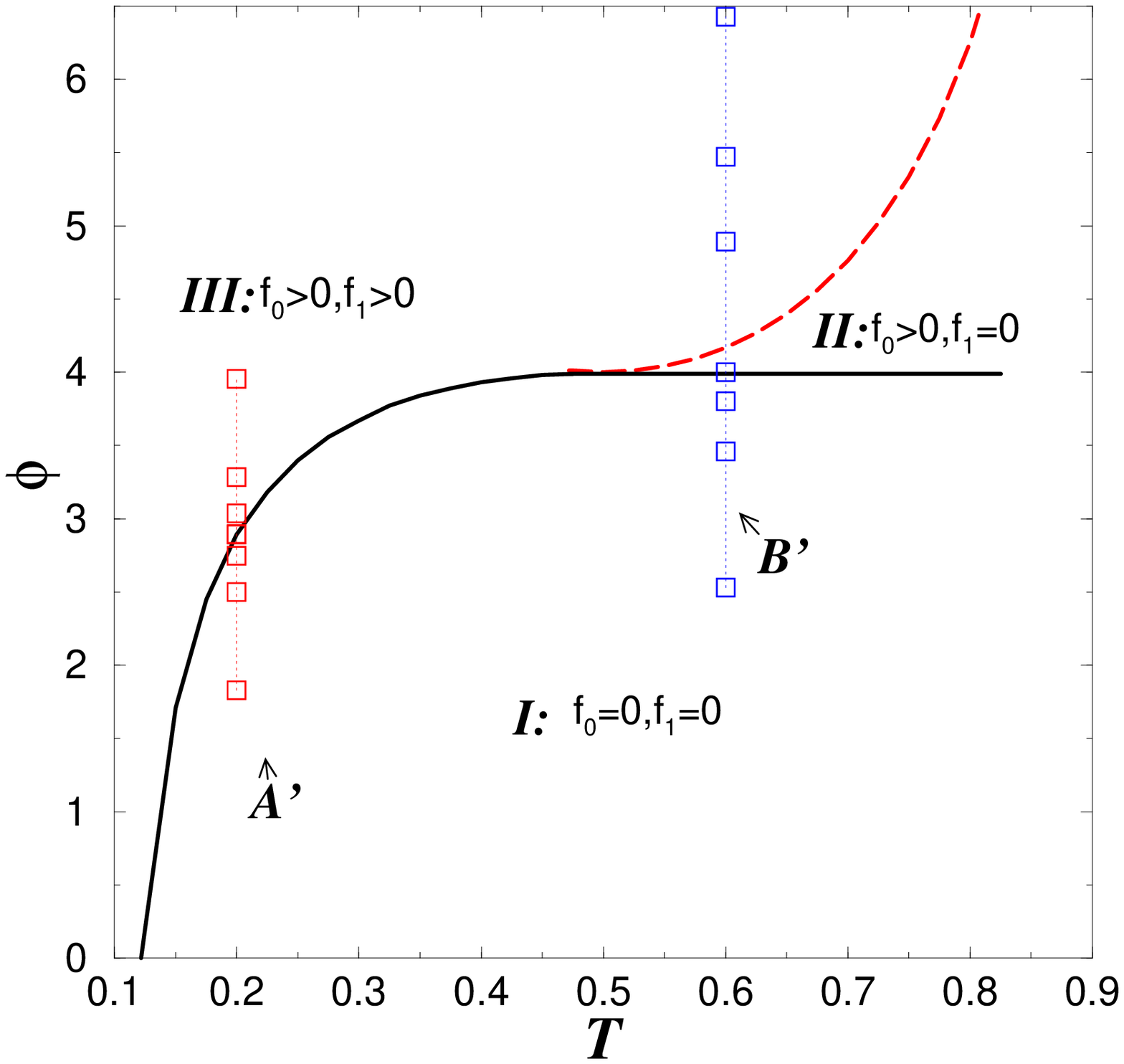}}}}
	\caption{Phase diagram for the glass transition between the phases I,
	II and III for the Bosse-Krieger
	model (\ref{eq43a}), (\ref{eq43}) and (\ref{eq44}) with $x_0=0.15$. 
	A' and B' denote paths along which we have investigated the dynamics}
\label{fig17}
\end{figure}

\begin{figure}
	\centerline{\rotatebox{-0}{\resizebox{7cm}{!}
	{\includegraphics{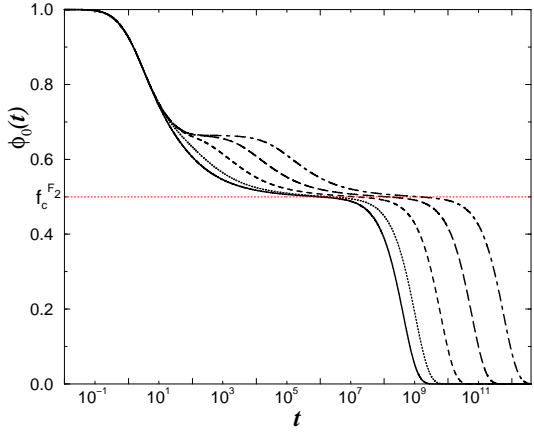}}}}
	\caption{$\phi_0(t)$ for the Bosse-Krieger
	model for $T=0.7$, $\phi=3.9995$, $x=0.15$, $\Omega_0=1$, $\nu_0=10$,
	$\nu_1=1$, and different $\Omega_1=10^{-n/2}$, $n=0,1,2,3,4$
	from left to right.
	$f_c^{F2}$ denotes the critical non-ergodicity parameter of the
	$F_2$-model}
\label{fig18}
\end{figure}

\begin{figure}
	\centerline{\rotatebox{-0}{\resizebox{9cm}{!}
	{\includegraphics{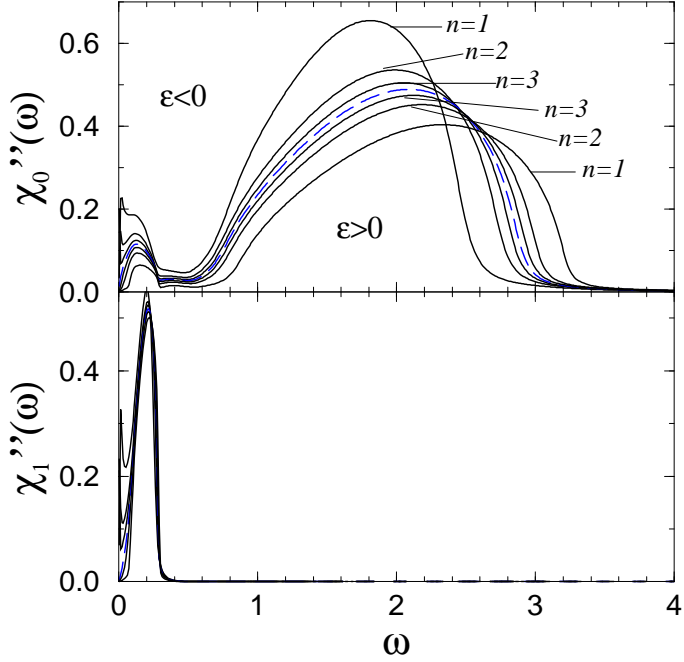}}}}
	\caption{$\chi_a''(\omega)$ for $\Omega_0=1$, $\Omega_1=1/10$,
	$\nu_0=\nu_1=0$, $x=0.15$  and $T=0.2$ along path A' for $\phi=
	(1+\varepsilon)\phi_c$, $\phi_c\simeq 2.8931318$,
	$\varepsilon=\pm e^{-n}$, $n=1,2,3$ (solid lines) and
        $\phi=\phi_c$ (dashed 
 	line). Because $\chi''_{1}(\omega)$
	does not vary much with $\phi$ we have not labelled the various curves
	with $n$} 
\label{fig20n}
\end{figure}

\begin{figure}	
\centerline{\rotatebox{-0}{\resizebox{9cm}{!}
	{\includegraphics{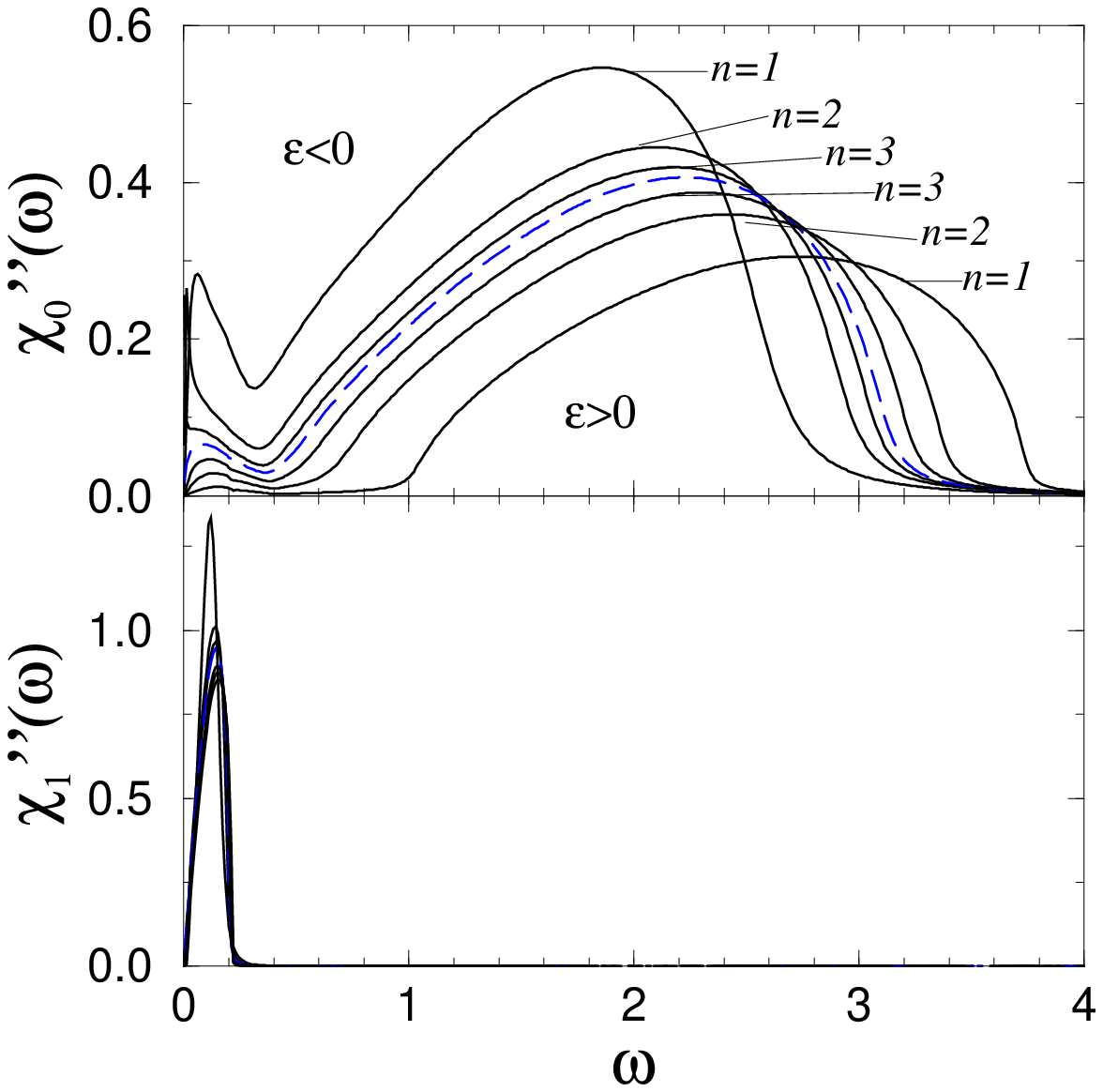}}}}
	\caption{$\chi_a''(\omega)$ for $\Omega_0=1$, $\Omega_1=1/10$,
	$\nu_0=\nu_1=0$, $x=0.15$ and $T=0.6$ along path B' for $\phi=
	(1+\varepsilon)\phi_c$, $\phi_c=4$, $\varepsilon=-e^{-n}$ for
	$\varepsilon<0$ and $\varepsilon=e^{-n/2}$ for
	$\varepsilon>0$, 
	$n=1,2,3$ (solid lines) and $\phi=\phi_c$ (dashed line). Because
	$\chi''_{1}(\omega)$ 
	does not vary much with $\phi$ we have not labelled the various curves
	with $n$} 
\label{fig21n}
\end{figure}

\begin{figure}
	\centerline{\rotatebox{-0}{\resizebox{9cm}{!}
	{\includegraphics{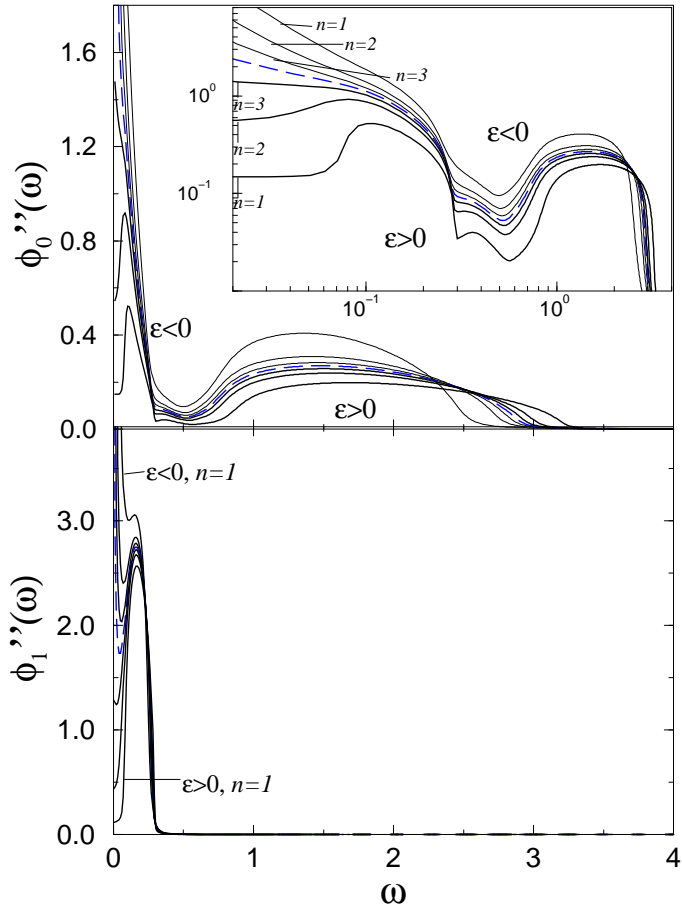}}}}
	\caption{$\phi_a''(\omega)$ for $\Omega_0=1$, $\Omega_1=1/10$,
	$\nu_0=\nu_1=0$, $x=0.15$  and $T=0.2$ along path A' for $\phi=
	(1+\varepsilon)\phi_c$, $\phi_c\simeq 2.8931318$,
	$\varepsilon=\pm e^{-n}$, $n=1,2,3$ (solid lines) and $\phi=\phi_c$
	(dashed line)} 
\label{fig22n}
\end{figure}

\begin{figure}
	\centerline{\rotatebox{-0}{\resizebox{9cm}{!}
	{\includegraphics{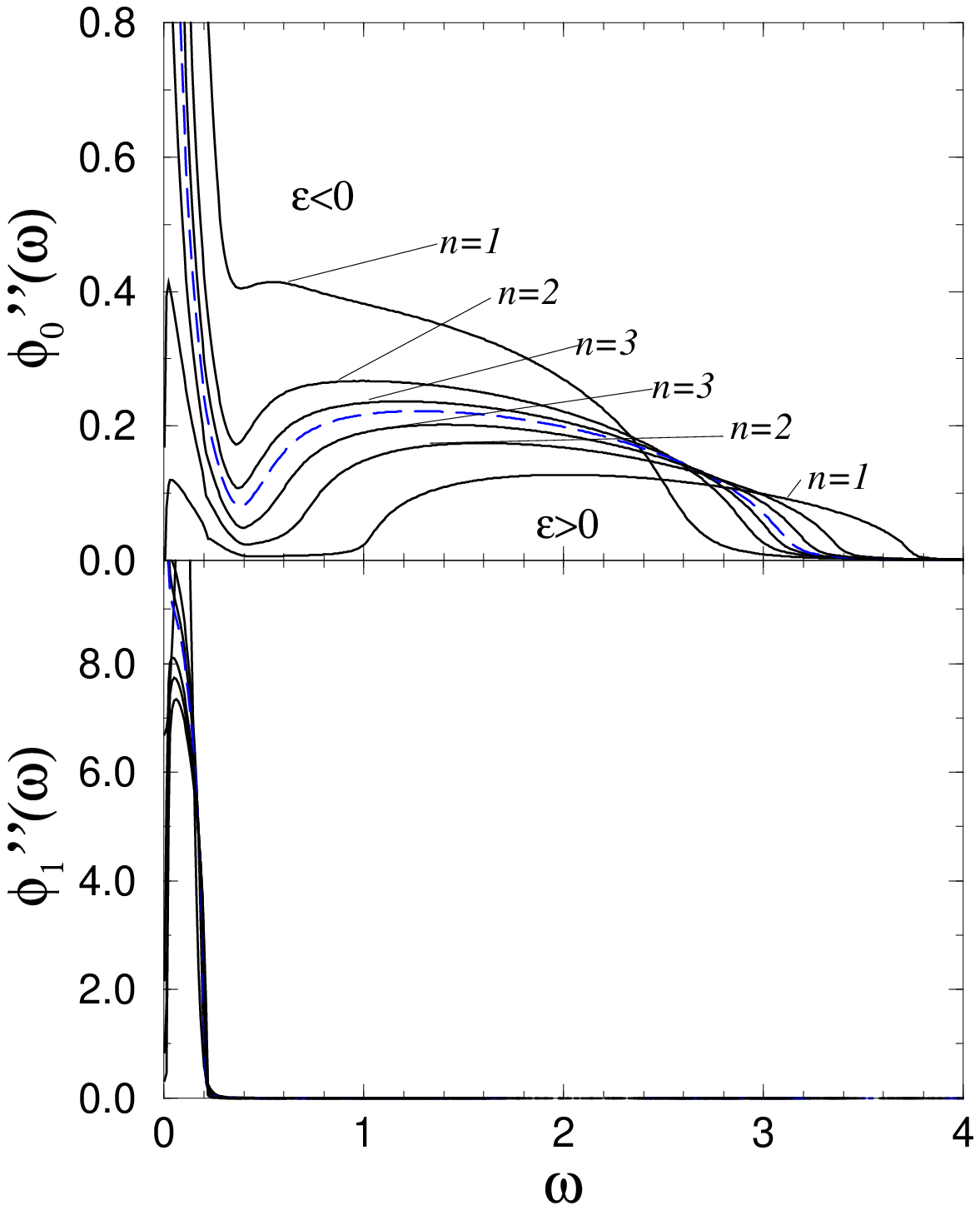}}}}
	\caption{$\phi_a''(\omega)$ for $\Omega_0=1$, $\Omega_1=1/10$,
	$\nu_0=\nu_1=0$, $x=0.15$ and $T=0.6$ along path B' for $\phi=
	(1+\varepsilon)\phi_c$, $\phi_c=4$, $\varepsilon=-e^{-n}$ for
	$\varepsilon<0$ and $\varepsilon=e^{-n/2}$ for
	$\varepsilon>0$, $n=1,2,3$ (solid lines) and $\phi=\phi_c$ (dashed
	line). Because $\phi''_{1}(\omega)$
	does not vary much with $\phi$ we have not labelled the various curves
	with $n$} 
\label{fig23n}
\end{figure}

\begin{figure}
	\centerline{\rotatebox{-0}{\resizebox{9cm}{!}
	{\includegraphics{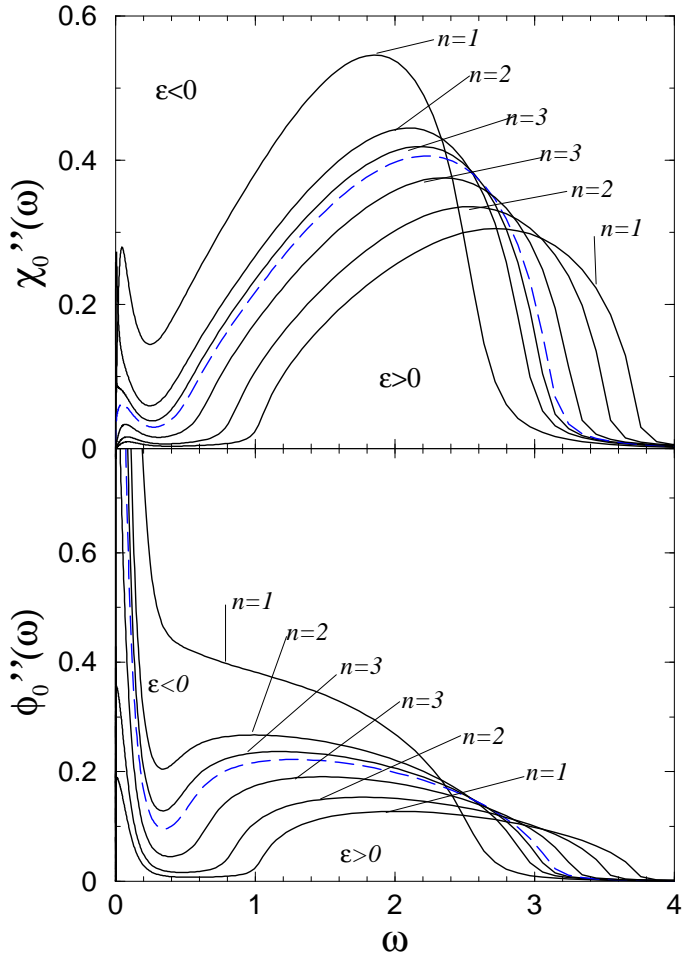}}}}
	\caption{$\chi_0''(\omega)$ and $\phi_0''(\omega)$ 
	for a reduced  Bosse-Krieger
	model $\Omega_0=1$, $\Omega_1=1/10$,
	$\nu_0=0$, $\nu_1=0.19$, $x=0.15$  and $T=0.6$  $\phi=
	(1+\varepsilon)\phi_c$, $\phi_c=4$, $\varepsilon=-e^{-n}$, for
	$\varepsilon<0$ and $\varepsilon=e^{-n/2}$ for
	$\varepsilon>0$,
	$n=1,2,3$ (solid lines) and $\phi=\phi_c$ (dashed line)}
\label{fig24n}
\end{figure}

A well-pronounced two-peak structure in the microscopic frequency regime can
also occur for $\chi_{10}''(q,\omega)$ (cf. Fig. \ref{fig7n}) deep in the glass,
whereas this does not happen for $\chi_{11}''(q.\omega)$
(cf. Fig. \ref{fig7n}) which is directly related to the dielectric loss
spectrum $\epsilon''(q,\omega)$.

Let us come back to the orientational peak. Figure \ref{fig17n}a shows that its
intensity varies linearly with temperature below about $T^*=0.13$ and for $
\phi_D=0.525$, i.e. for $T^*$ smaller than $T^*_c(\phi_D=0.525)\simeq 0.13$.
This fact, as well as the rather insensitivity of its position $\omega_{op}$
on $q$ and its location about one decade below the high frequency peak is
reminiscent to the boson peak. Additionally the increase of $\omega_{op}$
with packing fraction is similar to the shift of the boson peak to higher
frequency by increasing the pressure \cite{sugai96,inamura99,jund00}.

To conclude we can say that we have found in the microscopic frequency
regime of the compressibility
spectrum and the corresponding correlation spectrum
an additional orientational low frequency peak. 
It originates from a 
\emph{localized} and \emph{non-propagating orientational} mode coupled to
the \emph{longitudinal acoustic} sound waves (translational modes). 
Its dynamical origin changes from a pure relaxational to a damped
oscillational type when crossing the glass transition from the liquid side.
Since
the orientational frequency has a gap at $q=0$ the orientational excitation
is an \emph{optical} mode. These characteristic features coincide with
several explanations for the existence of the boson peak \cite
{buchenau84,taraskin97b,guillot97,ramos97,fischer99} which may 
stress the role
of orientational degrees of freedom. 
It is also interesting that the intensity of the op in a certain temperature
regime (cf. Fig. \ref{fig17n}a) changes linear with $T^*$ which is also true
for the boson peak.
In addition we have found an
interesting correlation between the orientational peak phenomenon and a medium range
orientational order, as already suggested earlier for the boson peak \cite
{elliot92,sokolov92,sokolov93,foley95,fayos96}, although it has been shown
that this does not seem to hold for all glass formers \cite{boerjesson93}. 

For dipolar hard spheres this order is of
ferroelectric type and for hard ellipsoids related to a precursor of nematic
order. For SiO$_2$ it can occur due to strong orientational bond
interactions. As demonstrated in the present paper
this medium range orientational order may result in a
softening of a localized, optic orientational mode and may finally generate
an additional microscopic peak about one decade below 
the high frequency peak. 

In general it is not so easy to separate from experimental spectra obtained
from light or neutron scattering the contribution from the orientational
motion.
But this is not true for a MD-simulation.
Due to the availability of all the microscopic information it would be
desirable to explore in more detail by MD-simulations the role of the
orientational degrees of freedom for the glass transition itself and also for
the spectra in the microscopic regime, as it has already been done for the
sound propagation in liquid water \cite{sciortino94}.

Finally we mention that the microscopic spectral features we have found
for DHS can be qualitatively reproduced by use of a schematic model, the
Bosse-Krieger model. \newline

ACKNOWLEDGMENTS
We are grateful to W.~G\"otze for his careful reading of this manuscript and
many valuable comments, to M.~Sperl and Th.~Voigtmann for their discussion with
respect to the accuracy of our numerical results and to P.~A.~Madden for the
discussion on orientational modes in molecular liquids. The financial support
by Sonderforschungsbereich 262 is gratefully acknowledged, as well.

\begin{appendix}
\section{The numerically solved MCT equations}
Using Eq. (\ref{eq4}) we get the first equation of motion for the
generalized density-density correlation functions (in the $q$-frame).
\begin{equation}
\label{eqA1}
\dot{S}_{ll'm}(q,t) = 
-i \sum_{\alpha=R,T} q^{\alpha}_l(q) \phi^{j^\alpha \rho}_{ll'm}(q,t).
\end{equation}
with
\begin{equation}
\label{eqA1b}
\phi^{j^\alpha \rho}_{ll'm}(q,t):=\frac{1}{N}\langle j^{\alpha *}_{lm}(q,t)\,
\rho_{l'm'}(q,0)\rangle \delta_{mm'}
\end{equation}
the {\em density - current-density correlation function}.
Projecting on the tensorial densities (Eq. (\ref{eq1})) and 
the longitudinal translational and rotational current densities (Eq. (\ref{eq5}))
at once leads to
\begin{eqnarray}
\nonumber
& & \dot{\phi}^{j^\alpha \rho}_{ll'm}(q,t)  =  \\
\label{eqA2}
        & - &i \sum_{l_2=m}^{l_{co}} q^{\alpha}_l(q) \frac{k_BT}{I_\alpha}
        ({\bf S}^{-1}(q))_{ll_2m} S_{l_2l'm}(q,t) \\
\nonumber 
        & - &  \sum_{l_2=m}^{l_{co}} 
	\sum_{\alpha'=R,T} \frac{k_BT}{I_\alpha} \int_0^t dt'
        { M}^{\alpha \alpha '}_{ll_2m}(q,t-t') 
	\phi^{j^{\alpha'} \rho}_{l_2l'm}(q,t') 
\end{eqnarray}
with the cut-off value $l_{co}$ for $l$ (for DHS it is $l_{co}=1$, cf.
Eqs. (\ref{eq39a}), (\ref{eq39b})) and $I_\alpha$ defined in (\ref{eq10}).
To make use of the established numerical method to solve such equations
\cite{fuchs91b} it is necessary to introduce auxiliary functions
$\Phi$:
\begin{equation}
\label{eqA3}
\dot{\Phi}^{j^\alpha \rho}_{ll'm}(q,t):=q^{\alpha}_l(q) \phi^{j^\alpha \rho}_{ll'm}(q,t)
\end{equation}
As a consequence Eq. (\ref{eqA1}) reduces to
\begin{equation}
\label{eqA4}
S_{ll'm}(q,t)=
-i \sum_{\alpha=R,T}{\Phi}^{j^\alpha \rho}_{ll'm}(q,t)
\end{equation}
where ${\Phi}^{j^\alpha \rho}_{ll'm}(q,t)$ is determined only up to an
integration constant which does not influence the result for $S_{ll'm}(q,t)$.
Making use of 
(\ref{eqA3})
and taking the time derivative of Eq. 
(\ref{eqA2}) 
leads to a  set of equations of similar  structure as the MCT
equations for simple liquids \cite{goetze91,goetze92,cummins99,schilling94}.
The molecular MCT -- equations do not only couple all ({\em
discretized})  wave
vectors, as in simple liquids, but also the indices $\alpha = T,\, R$
for translation and rotation and the spherical indices $l \le l_{co}$
and $ -l \le m \le l$. 
The functions $\Phi$ are auxiliary  functions because their 
introduction is just for numerical purpose.
The integral in Eq. (\ref{eqA2}) transforms to 
\begin{equation}
\label{eqA5}
\int_0^t dt' { M}^{\alpha \alpha '}_{ll_2m}(q,t-t') 
	\dot{\Phi}^{j^{\alpha'} \rho}_{l_2l'm}(q,t').
\end{equation}
Now a time derivative is multiplied with the integration measure $d t$.
As a consequence the time step $\Delta t$ cancels at a discretized 
form of (\ref{eqA5}).
Using the decimation technique \cite{fuchs91b} $\Delta t$ becomes very large.
In this case the calculation of the integrals would become
unstable if $\Delta t$ would not cancel.

Since the auxiliary functions only occur as time derivatives in the original
equations the result is independent on the initial values of $\Phi^{j^\alpha
\rho}_{ll'm}(q,t=0)$. When solving the molecular MCT equations we have
chosen: 
\begin{subeqnarray}
\Phi^{j^T \rho}_{ll'm}(q,t=0)& = & S_{ll'm}(q),\, \\ 
\Phi^{j^R \rho}_{ll'm}(q,t=0) & = & 0
\\\nonumber
\text{ for }  l=0,\forall l' \text{ and} \\
\Phi^{j^T \rho}_{ll'm}(q,t=0)& = &\frac{1}{2} S_{ll'm}(q),\, \\ 
\Phi^{j^R \rho}_{ll'm}(q,t=0) & = & \frac{1}{2} S_{ll'm}(q) 
\label{eqA6}
\end{subeqnarray}

for $l>0 $ as initial values.
\end{appendix}





\end{document}